%% file: icde2008_maybms.tex
\documentclass[times, 10pt,twocolumn]{article} 
\usepackage{latex8}
\usepackage{times}

\usepackage{epsfig}
\usepackage{latexsym}
\usepackage{amsmath}
\usepackage{amssymb}
\usepackage{amsthm}
\usepackage{subfigure}
\usepackage{pstricks}
\usepackage{pst-node}
\usepackage{txfonts}
\usepackage[vlined,ruled]{algorithm2e}
\usepackage{pst-tree}

\def\punto{$\hspace*{\fill}\Box$}
\newcommand{\nop}[1]{}
\newcommand{\shortversion}[1]{}
\newcommand{\longversion}[1]{#1}

\def\lBrack{\lbrack\!\lbrack}
\def\rBrack{\rbrack\!\rbrack}
\newcommand{\Bracks}[1]{\lBrack#1\rBrack}
\newcommand{\bananabr}[1]{\Bracks{#1}}

\newtheorem{theorem}{Theorem}[section]

\newtheorem{example}[theorem]{Example}
\newtheorem{definition}[theorem]{Definition}
\newtheorem{proposition}[theorem]{Proposition}

\newtheorem{lemma}[theorem]{Lemma}

\newcommand{\bluebox}[1]{\colorbox{blue!5}{#1}}

\title{Fast and Simple Relational Processing of Uncertain Data}
\author{Lyublena Antova, Thomas Jansen, Christoph Koch, and Dan Olteanu \\[1ex]
Saarland University Database Group \\
Saarbr\"ucken, Germany \\[1ex]
{\large \{}lublena, jansen, koch, olteanu{\large \}}@infosys.uni-sb.de}
\date{}

\begin{document}

\maketitle

\begin{abstract}
This paper introduces U-relations, a succinct and purely relational
representation system for uncertain databases. U-relations support
attribute-level uncertainty using vertical partitioning.
If we consider positive relational algebra extended by an operation for
computing possible answers, a query on the logical level can be translated
into, and evaluated as, a single relational algebra query on the U-relation
representation. The translation scheme
essentially preserves the size of the query in terms of
number of operations and, in particular, number of joins.
Standard techniques employed in off-the-shelf
relational database management
systems are effective for optimizing and processing queries on U-relations.
In our experiments we show that query evaluation on U-relations
scales to large amounts of data with high degrees of uncertainty.
%
%We also define a normal form for U-relations called WSDNF and provide a
%normalization algorithm. U-relations in WSDNF are of
%interest in tasks such as data cleaning and computing certain answers.
\end{abstract}

\input{introduction}

\input{representation}
\input{queries}

\input{normalization}

\input{discussion}

\input{experiments}

\input{conclusion}

{\small
\bibliographystyle{abbrv}\vspace*{-1em}
\bibliography{bibtex}
}

\end{document}

%% file: introduction.tex
\vspace{-3mm}

\section{Introduction}

\vspace{-2mm}

Several recent works
\cite{dalvi04efficient, CSP2005, BDSHW2006, miller06clean, DS2007, AKO07WSD, AKO07iWSD}
aim at developing scalable representation systems and query processing
techniques for  large collections of uncertain data as they arise in
data cleaning, Web data management, and scientific databases.
Most of them are based on a possible worlds semantics, and for all of them
such a semantics can be conveniently defined.

Four desiderata for representation systems for
incomplete information appear important.

\smallskip

%\begin{itemize}
%\addtolength{\topsep}{-0.3ex}
%\addtolength{\labelsep}{-0.3ex}
%\addtolength{\itemsep}{-1ex}
%\item

\noindent
{\bf 1. Expressiveness}\/.
The representation should be closed under the application of
(relational algebra) queries and data cleaning algorithms
(which remove some possible worlds).
That is, the results of
% applying
such operations to the
represented data should be again representable within the
formalism.

\smallskip

%\item

\noindent
{\bf 2. Succinctness}\/.
It should be possible to represent large sets of alternative worlds
using fairly little space. 

\smallskip

%\item

\noindent {\bf 3. Efficient query evaluation}\/.  A trade-off is
required between the succinctness of a representation formalism and
the complexity of evaluating interesting queries.  This trade-off
follows from established theoretical results
\cite{AKG1991,Gra1984,AKO07iWSD}. However, while the formalisms in the
literature tend to differ in succinctness, several have
polynomial-time data complexity for (decision) problems such as tuple
possibility under {\em positive}\/ (but not full) relational algebra.
This includes v-tables \cite{IL1984, Gra1984}, uncertainty-lineage
databases (ULDBs) \cite{BDSHW2006}, and world-set decompositions
(WSDs) \cite{AKO07iWSD}.

%In \cite{AKO07ISQL}, a nonsuccinct representation system is
%shown on which certain query evaluation problems are efficiently solvable
%that are NP-hard on more succinct representation systems, cf.\
%\cite{AKO06gWSD-Full}.

\smallskip

%\item

\noindent
{\bf 4. Ease of use}\/ for developers and researchers
in the sense that the representation system can be easily put on top of
a relational DBMS. 
This in particular includes
that queries on the logical schema level can be translated down to, ideally,
relational algebra queries on the
representation relations and that this translation is simple and
easy to implement.
This goal is motivated by the availability and maturity of
existing relational database technology.
%\end{itemize}

\smallskip

An important aspect of a representation system
is whether it represents uncertainty at the {\em attribute-level}\/
or the {\em tuple-level}\/. Attribute-level representation
refers to the succinct representation of relations
in which two or more fields of the same tuple can independently take
alternative values (see also \cite{AKO07iWSD}).
Attribute-level representation of uncertainty
(as supported by c-tables \cite{IL1984} and WSDs) 
offers finer granularity of independence than tuple-level approaches
such as \cite{BDSHW2006, dalvi04efficient, miller06clean}. This
is useful in applications like data cleaning in which the values of
several fields of a single tuple can be independently uncertain.
For instance, the U.S.\ Census Bureau maintains relations
with dozens of columns ($>$ 50), most of which may require cleaning
\cite{AKO07WSD}.

\medskip

\input{example}

\noindent
{\bf U-relations}.
In this paper, we develop and study {\em U-rela\-tions},
a representation system that we introduce with the following example.

\smallskip

\begin{example}\em
Let us assume that an aerial photograph of a battlefield
shows four vehicles at distinct positions 1 to 4. The resolution of the image
does not allow for the identification of vehicle types,
but we can draw certain conclusions from earlier reconnaissance and a
calculation of the maximum distance each vehicle may have covered since.
Say we know that vehicle
1 is (a) a friendly tank. Vehicles 2 and 3 are (b) a friendly
transport and (c) an
enemy tank, but we do not know which one is which. Nothing is known about
vehicle 4.
Figure~\ref{fig:u-rel}a shows a schematic drawing of how this scenario
can arise.
%, assuming previous reconnaissance
%has identified three vehicles (a), (b), and (c).
%
Only 1 is in the range of (a); 2 and 3 are in
the ranges of (b) and (c); and position 4 is near the border of the photograph
but outside the ranges of (a), (b), and (c), so this vehicle must have newly
moved onto the map.

We want to model this by an uncertain database of schema
$R$(Id, Coord, Type, Faction), representing the ids (1--4),
coordinate positions,
types, and factions of the vehicles on the map. Let us assume there are only
two vehicle types (tank or transport) and two factions (friend or enemy).
Then there are eight possible worlds. We obtain one by
taking three choices -- answering the following questions:
%\begin{enumerate}
%\item
Has the friendly transport (b) now become vehicle 2
($x \mapsto 1$) or 3 ($x \mapsto 2$)?
%
%\item
Is vehicle 4 a tank ($y \mapsto 1$) or a transport ($y \mapsto 2$)?
%
%\item
Is vehicle 4 friendly ($z \mapsto 1$) or an enemy ($z \mapsto 2$)?
%\end{enumerate}
Thus the uncertainty can be naturally
modelled using three variables $x, y, z$ that
each can independently take one of two values.

We model this scenario by the U-relational database shown in
Figure~\ref{fig:u-rel}b. We use vertical partitioning (cf.\ e.g.\
\cite{Bat1979,SAB+2005}) to achieve
attribute-level representation. $R$ is represented using four
U-relations, one for each column of $R$. The U-relation for the
coordinate positions (which are all certain) is not shown
since we do not want to use it subsequently, but of course, conceptually,
coordinate positions are an important feature of the example and have to be
part of the schema. In addition there is a relation $W$
which defines the possible values the three variables can take.

We can compute a vertical decomposition of one world 
given by a valuation $\theta$ of the variables $x,y,z$ by
(*) removing all the tuples from
the U-relations whose $D$ columns
contain assignments that are inconsistent
with $\theta$
(For example, if
$\theta = \{ x \mapsto 1, y \mapsto 1, z \mapsto 1 \}$ then we remove the
third and fifth  tuples of $U_1$ and the fifth tuples of $U_2$ and $U_3$.)
and then
(*) projecting the $D$ columns away.
Of course we can resolve the vertical partitioning
by joining the decomposed relations on the tuple id columns $T_R$.
\punto
\end{example}

%combines the advantages of ULDBs and WSDs at representing
%uncertain data.\footnote{We do not attempt to capture lineage as done in
%  ULDBs, though, as we consider it an orthogonal issue for which the
%decision what expressiveness is required in real applications (and achievable)
%is still unsettled \cite{BunemanICDT01}.}

%\smallskip

\noindent
U-relations have the following properties:
\begin{itemize}
\addtolength{\topsep}{-0.3ex}
\addtolength{\labelsep}{-0.3ex}
\addtolength{\itemsep}{-1ex}
\item
{\bf Expressiveness}:
U-relations are {\em complete}\/
for finite sets of possible worlds, that is, they
allow for the representation of any finite world-set.

\item
{\bf Succinctness}:
U-relations represent uncertainty on the attribute level.
Even though they allow for more efficient query evaluation, U-relations
are, as we show, exponentially more succinct than ULDBs and WSDs. That is,
there are (relevant) world-sets that necessarily take exponentially more space
to represent by ULDBs or WSDs than by U-relations.

\item {\bf Leveraging RDBMS technology}: U-relations allow for a large
  class of queries (positive relational algebra extended by the
  operation ``possible'') to be processed {\em using relational
    algebra only}\/, and thus efficiently in the size of the data.
  Our approach is the first so far to achieve this for the above-named
  query language. Indeed, this not only settles that there is a
  succinct and complete {\em attribute-level}\/ representation for which the
  so-called tuple Q-possibility problem
  for positive relational algebra is in polynomial time (previously open
  \cite{AKO07iWSD}) but puts a rich body of research results and technology at
  our disposal for building uncertain database systems.

  This makes U-relations the most efficient and
  scalable approach to managing uncertain databases to date.

\item
{\bf Parsimonious translation}:
The translation from relational algebra expressions on the
logical schema level to query plans on the physical representations replaces
a selection by a selection, a projection by a projection, a join by a join
(however, with a more intricate join condition), and a ``possible'' operation
by a projection. We have observed that state-of-the-art RDBMS do well
at finding efficient query plans for such physical-level queries.
\end{itemize}

\noindent
{\bf Ease of use:}
A main strength of U-relations is their simplicity and low ``cost of
ownership'':
\begin{itemize}
\addtolength{\itemsep}{-1ex}
\item
The representation system is purely relational and in close
analogy with relational representation schemes for vertically decomposed data.
Apart from the column
store relations that represent the actual data, there
is only a single auxiliary relation $W$
% that allows us to tell whether
%the empty world is possible
(which we need for computing certain answers, but not for possible answers).

\item
Query evaluation can be fully expressed in relational algebra. The
translation is quite simple and can even be done by hand,
at least for moderately-sized queries.

\item
The query plans obtained by our
translation scheme are usually handled well by the query optimizers of
off-the-shelf relational DBMS, so the implementation of special operators
and optimizer extensions is not strictly needed for acceptable performance.
\end{itemize}

Thus U-relations are not only suited as 
a representation system for dedicated uncertain database implementations such
as MayBMS \cite{AKO07WSD}, but are
also relevant to ``casual users'' of representation systems for
uncertain data, such as
researchers in data cleaning and data integration who want to store and
query uncertain data without great effort.

Apart from those implicitly mentioned above, we make the following further
contributions in this paper.

\begin{itemize}
\addtolength{\itemsep}{-1ex}
\item
We study algebraic query optimization and present equivalences
that hold on vertically decomposed representations. We address query
optimization using them in the context of managing uncertainty with
U-relations.

\item
We present an algorithm for normalizing a U-relational representation
obtained from a query.
%
%We introduce a normal form for U-relations, WSDNF,  and provide a
%normalization algorithm. U-relations in WSDNF
%directly correspond to World-set Decompositions (WSDs),
%a simple representation system for incomplete
%information that achieves succinctness using relational product decomposition
%\cite{AKO07WSD, AKO07iWSD}. WSDs have a number of nice properties, such as
%offering a convenient model for data cleaning.
%
Normalized U-relational databases yield a conceptually simple
algorithm for computing the certain answers of queries.
In particular, certain answer tuples on normalized
tuple-level representations can be computed using relational algebra only,
which is not true in general for previous representation systems.

\item
We provide experimental evidence for the efficiency
and relevance of our approach.
\end{itemize}

The structure of the paper is as follows.
Section~\ref{sec:u-relations} establishes
U-relations formally.
Section~\ref{sec:queries} presents our reduction from queries on the logical
level to relational algebra on the level of U-relations and addresses
algebraic query evaluation.
Section~\ref{sec:normalization} presents the normalization algorithm.
Section~\ref{sec:discussion} discusses
the relationship between U-relations, WSDs and ULDBs and argues that
U-relations combine the advantages of the other two formalisms without
sharing their drawbacks.
In Section~\ref{sec:experiments}, we report on our experiments with
U-relations. We conclude with Section~\ref{sec:conclusion}.

%% file: example.tex
\begin{figure*}
\begin{small}
\epsfig{file=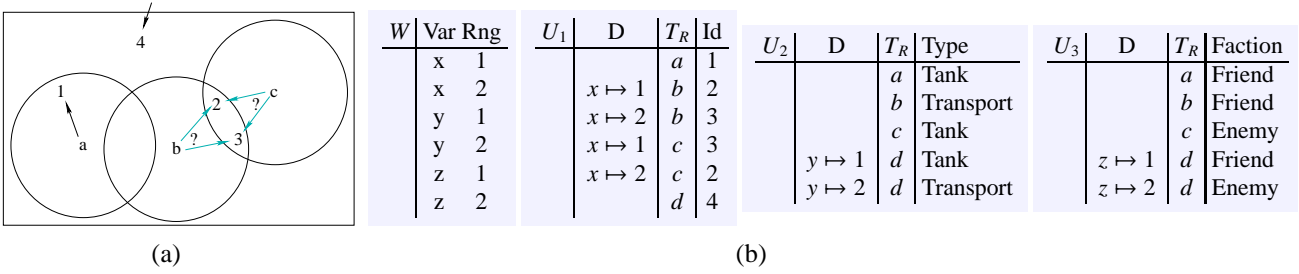, height=3cm}

\vspace{-2.9cm}
\hspace{4.8cm}
\bluebox{
\begin{tabular}{@{~}c@{~}|@{~}c@{~}c@{~}}
$W$ & Var & Rng \\
\hline
& x & 1 \\
& x & 2 \\
& y & 1 \\
& y & 2 \\
& z & 1 \\
& z & 2
\end{tabular}}
\bluebox{
\begin{tabular}{@{~}c@{~}|@{~~}c@{~~}|@{~}c@{~}|@{~}c@{~}}
$U_1$ & D & $T_R$ & Id \\
\hline
&               & $a$ & 1 \\
& $x \mapsto 1$ & $b$ & 2 \\ 
& $x \mapsto 2$ & $b$ & 3 \\ 
& $x \mapsto 1$ & $c$ & 3 \\ 
& $x \mapsto 2$ & $c$ & 2 \\ 
&               & $d$ & 4
\end{tabular}}
\bluebox{
\begin{tabular}{@{~}c@{~}|@{~~}c@{~~}|@{~}c@{~}|@{~}l@{~}}
$U_2$ & D & $T_R$ & Type \\
\hline
&               & $a$ & Tank \\
&               & $b$ & Transport \\ 
&               & $c$ & Tank \\ 
& $y \mapsto 1$ & $d$ & Tank \\
& $y \mapsto 2$ & $d$ & Transport
\end{tabular}}
\bluebox{
\begin{tabular}{@{~}c@{~}|@{~~}c@{~~}|@{~}c@{~}|@{~}l@{~}}
$U_3$ & D & $T_R$ & Faction \\
\hline
&               & $a$ & Friend \\
&               & $b$ & Friend \\ 
&               & $c$ & Enemy \\ 
& $z \mapsto 1$ & $d$ & Friend \\
& $z \mapsto 2$ & $d$ & Enemy
\end{tabular}}
\end{small}

\medskip

\hspace{1.9cm} (a)
\hspace{7.2cm} (b)

\vspace{-.75em}

\caption{Map with moving vehicles (a) and U-relational database representation
of the possible worlds at the time the aerial photograph detecting vehicles
1,2,3,4  was taken (b).}
\label{fig:u-rel}
\vspace{-2em}
\end{figure*}

%%% Local Variables: 
%%% mode: latex
%%% TeX-master: "icde2008_maybms"
%%% End: 

%% file: representation.tex
\section{U-relational databases}
\label{sec:u-relations}

% U-relational databases represent finite sets of possible worlds, where
% each world consists of one relational database. They form a very
% succinct representation formalism for large world-sets due to explicit
% representation of (in)dependencies of field values across worlds and
% vertical relational decomposition.

We define \textit{world-sets} in close analogy to the case of
c-tables~\cite{IL1984}. Consider a finite set of variables over finite
domains.  A \textit{possible world} is represented by a total
valuation (or assignment) $f:$Var $\mapsto$ Rng of variables to
constants in their domains, and the world-set is represented by the
finite set of all total valuations\footnote{This is a generalization
  of world-set decompositions of~\cite{AKO07WSD}, where component ids
  are variables and local world ids are domain values.}.  We represent
relationally the variable set and the associated domains by a
\textit{world-table} over schema $W$(Var,Rng) such that $W$ consists
of all pairs $(x,v)$ of variables $x$ and values $v$ in the domain of
$x$.

\begin{example}\em
  The world-table $W$ in Figure~\ref{fig:u-rel} defines three
  variables $x, y, z$, whose common domain is $\{1,2\}$. The number of
  worlds defined by $W$ is $2\cdot 2\cdot 2 = 8$.\punto
\end{example}

Given a world-table $W$, a \textit{world-set descriptor} over $W$, or
ws-descriptor for short, is a valuation $\overline{d}$ such that its
graph is a subset of $W$. If $\overline{d}$ is a \textit{total}
valuation, then it represents one world. In our examples, to represent
the entire world-set we use an \textit{empty} ws-descriptor, as a
shortcut for a singleton ws-descriptor with a new variable with a
singleton domain.

%% We associate ws-descriptors to each possible value of a tuple field defined by
%% a U-relational database. This allows us to encode compactly the set of worlds
%% in which a tuple field has a particular value, and also to reason about the
%% (in)dependency of tuple fields. We additionally allow to define fields of the
%% same tuple separately by using vertical decomposition, as exemplified next.

%% \begin{example}\em
%%   Consider the U-relational database of Figure~\ref{fig:u-rel}
%%   consisting of the world-table $W$ and two U-relations $U_1$ and
%%   $U_2$. The first tuple of $U_1$ defines the value $1$ for the
%%   attribute CUST of the tuple $o_1$ od relation Ord in all worlds
%%   represented by the dependency vector $(c_2:1)$. Because $1$ is the
%%   only possible value for $c_2$, the tuple field Ord.$o_1.$CUST has
%%   value $1$ in all worlds. 

%%   The first tuple of $U_2$ defines the value 2003 for the attribute
%%   DATE of the same tuple $o_1$ of Ord. In this case, however, that
%%   value only occurs in a half of the possible worlds, namely those
%%   defined by $(c_1:1)$ and any other assignment for the remaining
%%   variables.\punto
%% \end{example}

%%%%%%%%%%%%%%%%%%%%%%%%%%%%%%%%%

We are now ready to define databases of U-relations.

\begin{definition}\label{def:u-rel}
\em
  A {\em U-relational database} for a world-set over schema
  $\Sigma = (R_1[\overline{A_1}], \dots, R_k[\overline{A_k}])$ is a tuple
\[
  (U_{1,1}, \ldots, U_{1,m_1}, \ldots, U_{k,1}, \ldots, U_{k,m_k}, W),
\]
where $W$ is a world-table and each relation $U_{i,j}$ has schema
$U_{i,j}[\overline{D}_{i,j};$ $\overline{T}_{R_i}; \overline{B_{i,j}}]$ such
that $\overline{D}_{i,j}$ defines ws-descriptors over $W$,
$\overline{T}_{R_i}$ defines tuple ids, and $\overline{B_{i,1}} \cup \dots
\cup \overline{B_{i,m_i}} = \overline{A}_i$.
\end{definition}

A ws-descriptor $\{c_1\mapsto l_1,\ldots,c_k\mapsto l_k\}$ is
relationally encoded in $\pi_{\overline{D}_{i,j}}(U_{i,j})$ of arity
$n\geq k$ as a tuple $(c_1\mapsto l_1,\ldots,c_k\mapsto
l_k,c_{k+1}\mapsto l_{k+1},\ldots,c_n\mapsto l_n)$, where each
$c_i\mapsto l_i$ is a $c_j\mapsto l_j$ for any $j$ and all $i$ with
$1\leq j\leq k<i\leq n$.

Although we speak of vertical partitioning, we do not require the value
columns of $U_{i,j}$ to disjointly partition the columns of $R_i$. Indeed,
overlap may be useful to speed up query evaluation, see e.g.\ \cite{SAB+2005}.
%
%% We will also use the notation $U^{R; \overline{A}}$ for a U-relation that
%% represents the columns $\overline{A}$ of relation $R$.

%% Let $f$ be a total valuation and $\overline{d} = (c_1\mapsto l_1, \dots,
%% c_n\mapsto l_n)$ be a ws-descriptor. We say that $\overline{d}$ {\em
%%   matches}\/ the world given by $f$ iff function $f$ extends $\overline{d}$
%% (i.e., for all $x$ on which $\overline{d}$ is defined,
%% $\overline{d}(x)=f(x)$).

We next define the semantics of a U-relational database. To obtain a possible
world we first choose a total valuation $f$ over $W$. We then process the
U-relations tuple by tuple. If the function $f$ extends\footnote{That is, for
  all $x$ on which $\overline{d}$ is defined, $\overline{d}(x)=f(x)$.} the
ws-descriptor $\overline{d}$ of a tuple of the form $(\overline{d},
\overline{t}, \overline{a})$ from a U-relation of schema $(\overline{D},
\overline{T}, \overline{A})$, we insert in that world the values
$\overline{a}$ into the $\overline{A}$-fields of the tuple with identifier
$\overline{t}$. In general this may leave some tuples partial in the end
(i.e., the values for some fields have not been provided.) These tuples are
removed from the world.

We require, for a U-relational database $(U_1, \ldots, U_n, W)$ to be
considered valid, that the representation does not provide several
contradictory values for a tuple field in the same world. Formally, we
require, for all $1\leq i,j\leq n$, and tuples $t_1\in
U_i[\overline{D}_i,\overline{T}_i,\overline{A}_i]$ and $t_2\in
U_j[\overline{D}_j,\overline{T}_j,\overline{A}_j]$ such that $U_i$ and $U_j$
are vertical partitions of the same relation, that if there is a world that
extends both $t_1.\overline{D}_i$ and $t_2.\overline{D}_j$, then for all $A
\in (\overline{A}_i\cap\overline{A}_j)$, $t_1.A=t_2.A$ must hold.

\begin{example}\em
Suppose there are two U-relations with schemata
$U_1[\overline{D_1}; T_R; A, B]$ and
$U_2[\overline{D_2}; T_R; B, C]$ that jointly represent columns
$A$, $B$, and $C$ of a relation $R$.
Assume tuples $(c_1, 1, t_1, a, b) \in U_1$ and
$(c_2, 2, t_1, b', c) \in U_2$.
Then $U_1$ and $U_2$ cannot form part of
a valid U-relational data\-base because
there would be a world with $c_1 \mapsto 1, c_2 \mapsto 2$ in which
the tuple from $U_1$ requires field $t_1.B$ to take value $b$ while the
tuple from $U_2$ requires the same field to take value $b$'.\punto
\end{example}

A salient property of U-relational databases is that they form a {\em complete
  representation system} for finite world-sets.

\begin{theorem}
  Any finite set of worlds can be represented as a U-relational
  database.
\end{theorem}

%%% Local Variables: 
%%% mode: latex
%%% TeX-master: "icde2008_maybms"
%%% End: 

%% file: queries.tex
\section{Query Processing}
\label{sec:queries}

The semantics of a query $Q$ on a world-set is to evaluate $Q$ in each
world.  For complete representation systems like U-relational
databases, there is an equivalent, more efficient
approach~\cite{IL1984}: Translate $Q$ into a query $\hat{Q}$ such that
the evaluation of $\hat{Q}$ on a U-relational encoding of the
world-set produces the U-relational encoding of the answer to $Q$.
\vspace*{.5em}

{\noindent\bf Queries on vertical decompositions.}  U-relations rely
essentially on vertical decomposition for succinct (attribute-level)
representation of uncertainty. To evaluate a query, we first need to
reconstruct relations from vertical decompositions by (1) joining two
partitions on the common tuple id attributes and (2) discarding the
combinations that yield inconsistent ws-descriptors. We call this
operation \textit{merge} and give its precise definition in
Figure~\ref{fig:microops}, where the two above conditions are defined
by $\alpha$ and $\psi$, respectively.

\begin{example}\label{ex:query-2ndlayer}
  \em Consider the U-relational database of Figure~\ref{fig:u-rel}.
  The query
  $\sigma_{\mathrm{Faction='Enemy'}\wedge\mathrm{Type='Tank'}}(R)$
  lists the enemy tanks on the map. To answer this query, we need to
  \textit{merge} the necessary partitions of $R$ and obtain a new
  query with $merge(\pi_{\mathrm{Faction}}(R),\pi_{\mathrm{Type}}(R))$
  in the place of $R$. \punto
\end{example}

Our query evaluation approach can take full advantage of query
evaluation and optimization techniques on vertical partitions. First,
it does not require to reconstruct the entire relations involved in
the query, but rather only the necessary vertical partitions. Second,
necessary partitions can be flexibly merged in during query
evaluation. Thus early and late tuple materialization~\cite{SAB+2005}
carry over naturally to our framework. For this, our \textit{merge}
operator allows to merge two partitions not only if they are given in
their original form, but also if they have been modified by queries.

The first advantage only holds for so-called {\em reduced}
U-relational databases, which do not have tuples that cannot be
completed in any world. That is, each tuple of a reduced U-relation
can always be completed to an actual tuple in a world. The advantage
becomes evident even for a simple projection query. Consider a reduced
database containing a U-relation $U$ defining the $A$ attribute of
$R$. To evaluate $\pi_{A}(R)$ we do not need to merge in all
U-relations defining the attributes of $R$ and later project on $A$.
Instead, the answer is simply $U$. In the following, we assume that
the input database is always reduced. As we will discuss next, our
query evaluation technique always produces reduced U-relations for
reduced input U-relational databases.

\begin{example}
\label{ex:wsdnf}
\em Consider the following non-reduced database of two U-rela\-tions:
\begin{center}
  \begin{small}    
    \begin{tabular}{c}

    \bluebox{ 
   \begin{tabular}{@{\extracolsep{0.4cm}}c|@{\extracolsep{0.2cm}}c|@{\extracolsep{0.2cm}}c|@{
\extracolsep{0.2cm}}c}
        $U_1$ & $D$      & $T$ & A\\
        \hline
        & $c_1\mapsto 1$ & $t_1$ & $a_1$\\
        & $c_2\mapsto 1$ & $t_2$ & $a_2$\\
      \end{tabular}
      }
      \hspace{0.3cm}
      \bluebox{
      \begin{tabular}{@{\extracolsep{0.4cm}}c|@{\extracolsep{0.2cm}}c|@{\extracolsep{0.2cm}}c|@{\extracolsep{0.2cm}}c}
        $U_2$ & $D$      & $T$ & B\\
        \hline
        & $c_1\mapsto 1$ & $t_1$ & $b_1$\\
        & $c_1\mapsto 2$ & $t_1$ & $b_2$\\
      \end{tabular}
      }
    \end{tabular}
  \end{small}
\end{center}
In each U-relation the second tuple cannot find a partner in the other
U-relation with which a complete tuple (with both attributes A and B) can be
formed. If these second tuples are removed, the database is reduced.  \punto
\end{example}

% TO CK: What is the purpose of this remark?
%% \begin{remark}\em
%%   When talking about reduced databases, it makes sense to assume that the
%%   U-relations representing a relation $R$ do not overlap in the value columns.
%%   (We do not want to reduce databases that have intentional redundancy.)\punto
%% \end{remark}

We can always reduce a U-relational database as follows: We filter
each U-relation using semijoins with each of the other U-relations
representing data of the same relation $R_i$. The semijoin conditions
are the $\alpha$ and $\psi$-conditions.

\begin{proposition}
  Given a schema $\Sigma$, there is a relational algebra query that
  reduces a U-relational database over $\Sigma$.
\end{proposition}

\begin{figure}
    \begin{align}
      & merge(\pi_{\overline{X}}(R), \pi_{\overline{A}-\overline{X}}(R)) = R,
      \hspace*{2em} \mbox{ where } \overline{A} = \textbf{sch}(R)\\
      & merge(R, S) = merge(S, R)\\
      & merge(merge(R, S), T) = merge(R, merge(S, T))\\
      & \sigma_{\phi(\overline{X})}(merge(R, S)) = merge(\sigma_{\phi(\overline{X})}(R), S)\\
      & \hspace*{8em} \mbox{ where } \overline{X} \subseteq \textbf{sch}(R) \nonumber\\
      & merge(R, S) \bowtie_{\phi(\overline{X}, \overline{Y})} T =
      merge(R \bowtie_{\phi(\overline{X}, \overline{Y})} T, S)\\
      & \hspace*{8em} \mbox{ where } \overline{X} \cup \overline{Y} \subseteq \textbf{sch}(R) \cup \textbf{sch}(T)\nonumber\\
      & \pi_{\overline{X}}(merge(R, S)) = merge(\pi_{\overline{X} \cap
        \overline{A}}(R), \pi_{\overline{X} \cap \overline{B}} (S))\\
      & \hspace*{8em} \mbox{ where } \textbf{sch}(R) = \overline{A}, \textbf{sch}(S) = \overline{B}\nonumber
    \end{align}

\vspace{-1em}

\caption{Algebraic equivalences for relational algebra queries with merge operator.}
\label{fig:equiv}
\vspace{-2em}
\end{figure}

{\noindent\bf Algebraic equivalences.} Figure~\ref{fig:equiv} gives
algebraic equivalences of relational algebra expressions with merge
operator on vertical decompositions: Merging is the reverse of
vertical partitioning, it is commutative and associative, it commutes
with selections, joins, and projections.

Standard heuristics known from classical query optimization for
relational algebra apply here as well. Intuitively, we usually push
down projections and selections and merge in U-relations as late as
possible. An interesting new case is the decision on join ordering
among an explicit join from the input query and a join due to merging:
If the merge is executed before the explicit join, it may reduce the
size of an input relation to join. We have seen in our experiments
that the standard selectivity-based cost measures employed by
relational database management systems do a good job, as long as the
queries remain reasonably small.

\begin{figure}[!t]
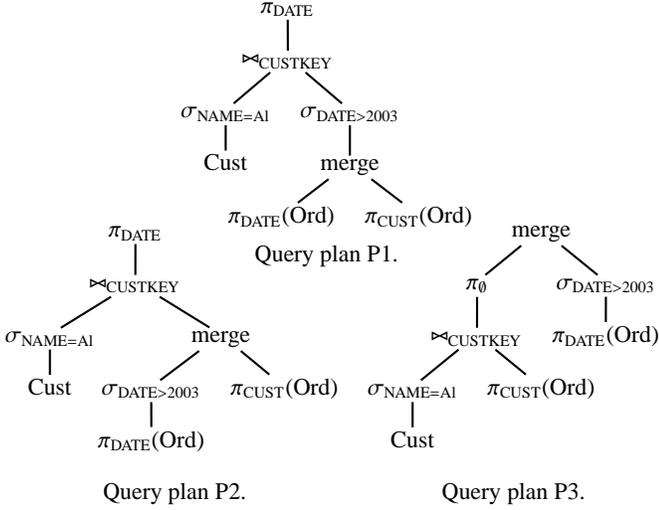

  \centering
  \begin{small}
                \pstree[levelsep=.7cm,treesep=.4cm,nodesep=.05cm]{\TR{$\pi_{\textrm{DATE}}$}}
                {
                        \pstree{\TR{$\bowtie_{\textrm{CUSTKEY}}$}}
                        {
                                \pstree{\TR{$\sigma_{\textrm{NAME=Al}}$}}
                                {
                                        \TR{Cust}
                                }
                                \pstree{\TR{$\sigma_{\textrm{DATE$>$2003}}$}}
                                {
                                        \pstree{\TR{merge}}{
                                                \TR{$\pi_{\textrm{DATE}}(\textrm{Ord})$}
                                                \TR{$\pi_{\textrm{CUST}}(\textrm{Ord})$}
                                        }
                                }
                        }
                }

        \medskip

        Query plan P1.

\vspace{-0.8cm}

\hspace*{-1em}%
  \begin{tabular}{c@{\hspace*{1em}}c}
        \pstree[levelsep=.7cm,treesep=.4cm,nodesep=.05cm]{\TR{$\pi_{\textrm{DATE}}$}}
        {
                \pstree{\TR{$\bowtie_{\textrm{CUSTKEY}}$}}
                {
                        \pstree{\TR{$\sigma_{\textrm{NAME=Al}}$}}
                        {
                                \TR{Cust}
                        }
                        \pstree{\TR{merge}}{
                        {
                                \pstree{\TR{$\sigma_{\textrm{DATE$>$2003}}$}}
                                {
                                        \TR{$\pi_{\textrm{DATE}}(\textrm{Ord})$}}
                                        \TR{$\pi_{\textrm{CUST}}(\textrm{Ord})$}
                                }
                        }
                }
        }
        &
        \pstree[levelsep=.7cm,treesep=.4cm,nodesep=.05cm]{\TR{merge}}
        {
                \pstree{\TR{$\pi_{\emptyset}$}}
                {
                        \pstree{\TR{$\bowtie_{\textrm{CUSTKEY}}$}}
                        {
                                \pstree{\TR{$\sigma_{\textrm{NAME=Al}}$}}
                                {
                                        \TR{Cust}
                                }
                                \TR{$\pi_{\textrm{CUST}}(\textrm{Ord})$}
                        }
                }
                \pstree{\TR{$\sigma_{\textrm{DATE$>$2003}}$}}
                {
                  \TR{$\pi_{\textrm{DATE}}(\textrm{Ord})$}
                }
        }
        \\
        \mbox{}\\
        Query plan P2.
        &
        Query plan P3.
        \end{tabular}
        \end{small}

\vspace{-1em}

\caption{Three equivalent query plans.}
\label{fig:qopt}
\vspace{-1.5em}
\end{figure}

\begin{example}\label{ex:heuristics}
  \em Consider a U-relational database ${\cal U}$ that represents a
  set of possible worlds over two TPC-H relations Ord and Cust (short
  for Order and Customer, respectively)~\cite{tpch2006}. ${\cal U}$
  has one U-relation for each attribute of the two relations, of which
  we only list DATE and CUSTKEY for \textrm{Ord}, and NAME and CUSTKEY
  for \textrm{Cust}.  The following query finds all dates of orders
  placed by Al after 2003:
  \begin{align*}
    \pi_{\mathrm{DATE}} (& \sigma_{\mathrm{NAME =
        'Al'}}(\mathrm{Cust}) \bowtie_{\mathrm{CUSTKEY}}
    \sigma_{\mathrm{DATE > 2003}}(\mathrm{Ord}))
  \end{align*}
  Figure~\ref{fig:qopt} shows three possible plans P1, P2, and P3
  using operators on vertical decompositions. The na\"{i}ve plan P1
  first reconstructs Ord from its two partitions then applies the
  selection and the join with Cust. In P2 and P3 the merge operator is
  pushed up in the plans, first immediately above the selection (P2),
  and then above the join operator (P3). Among the three plans, P1 is
  clearly the least efficient. However, without statistics about the
  data, one cannot tell which of P2 and P3 should be preferred. If
  DATE$>$2003 is very selective, then merging immediately thereafter
  as in P2 will lead to filtering of tuples from
  $\pi_{\mathrm{CUSTKEY}}(\mathrm{Ord})$ and thus fewer tuples will be
  processed by the join. Is this not the case, then first merging only
  increases the number and size of the tuples that have to be
  processed by the join. Also, in P3 all value attributes except of
  DATE are projected away after the join as they are not needed for
  the final result.  \punto
\end{example}

%%%%%%%%%%%%%%%%%%%%%%%%%%%%%%%%%%%%%%%%%%%%%%%%%%%%%%%

{\noindent\bf Queries on U-relations.}  Figure~\ref{fig:microops}
gives the function $\bananabr{\cdot}$ that translates positive
relational algebra queries with \textit{poss} and \textit{merge}
operators into relational algebra queries on U-relational databases.

The poss operator applied on a U-relation $U$ closes the possible worlds
semantics by computing the set of tuples possible in $U$. It thus translates
to a simple projection on the value attributes of $U$. The result of a
projection is a U-relation whose value attributes are those from the
projection list (thus the input ws-descriptors and tuple ids are preserved).
Selections apply conditions on the value attributes.

The merge operator that reconstructs a relation from its vertical partitions
was already explained. Similarly to the merge, the
join uses the $\psi$-condition to discard tuple combinations with inconsistent ws-descriptors.
Figure~\ref{fig:microops}
gives the translation in case $U_1$ and $U_2$ do not contain
partitions of the same relation. For the case of self-joins
we require aliases for the copies of the relation involved in it such that they
do not have common tuple id attributes.

The union of $U_1$ and $U_2$ like the ones from Figure~\ref{fig:microops} is
sketched next. We assume that $\overline{A}_1=\overline{A}_2$,
$\overline{T}_1\cap\overline{T}_2=\emptyset$, and the tuples of
different relations have different ids. To bring $U_1$ and $U_2$ to
the same schema, we first ensure ws-descriptors of the same size by
pumping in the smaller ws-descriptors already contained variable
assignments, and add new (empty) columns $\overline{T}_2$ to $U_1$ and
$\overline{T}_1$ to $U_2$. We then perform the standard union.

\begin{figure}
\begin{center}
\framebox{
  \parbox{6cm}{
    \hspace*{-.1em}\vspace*{-1em}\begin{align*}
      \mbox{Let } U_1 &:= \bananabr{Q_1} \mbox{ with schema } [\overline{D}_1, \overline{T}_1, \overline{A}_1],\hspace*{3em}\\
      U_2 &:= \bananabr{Q_2} \mbox{ with schema } [\overline{D}_2, \overline{T}_2, \overline{A}_2],\\
      \alpha &:= \underset{T\in\overline{T}_1\cap\overline{T}_2}{\bigwedge}(U_1.T = U_2.T),\\
      \psi &:= \underset{D'\in U_1.\overline{D}_1, D''\in U_2.\overline{D}_2}{\bigwedge(D'.\mbox{Var}=D''}.\mbox{Var} \Rightarrow D'.\mbox{Rng}=D''.\mbox{Rng}).
    \end{align*}\vspace*{-1.5em}
    \begin{align*}
      \bananabr{poss(Q_1)} &:= \pi_{\overline A_1}(U_1)\\
      \bananabr{\pi_{\overline{X}} (Q_1)} &:= \pi_{\overline{D}_1,\overline{T}_1, \overline{X}} (U_1),
      \hspace{1em} \mbox{where } \overline{X} \subseteq \overline{A}_1 \\
      \bananabr{\sigma_{\phi} (Q_1)} &:= \sigma_{\phi} (U_1),
      \hspace{3em} \mbox{where } \phi \mbox{ on } \overline{A}_1 \\
      \bananabr{Q_1 \bowtie_{\phi} Q_2} &:=
      \pi_{\overline{D}_1, \overline{D}_2, \overline{T}_1, \overline{T}_2, \overline{A},\overline{B}}
      (U_1 \bowtie_{\phi \land \psi} U_2), \hspace*{3em}\\
      &\hspace*{7.5em}\mbox{ where } \overline{T}_1 \cap \overline{T}_2 = \emptyset\\
      \bananabr{merge(Q_1, Q_2)} &:= 
      \pi_{\overline{D}_1,\overline{D}_2,\overline{T}_1 \cup\overline{T}_2,
        \overline{A},\overline{B}}(U_1 \bowtie_{\alpha \land \psi} U_2)
    \end{align*}
  }
}
\end{center}

% %%%%
% %        $\bananabr{\pi_{\overline A}(Q_1 \times Q_2)} :=
% %        \bananabr{Q_1} \times
% %        \pi_{\overline{T}_2}(\bananabr{Q_2})$ \\
% %        \\
% %%%%

\vspace{-1em}

\caption{Translation of queries with merge into queries on U-relations.}
\label{fig:microops}
\vspace{-2em}
\end{figure}

From our translation $\bananabr{\cdot}$ it immediately follows that

\begin{theorem}
  Positive relational algebra queries extended with the possible
  operator can be evaluated on U-relational databases using relational
  algebra only.
\end{theorem}

\begin{example}\label{ex:query}
  \em Recall the U-relational database of Figure~\ref{fig:u-rel}
  storing information about moving vehicles. Consider a query asking
  for ids of enemy tanks:
\[
  S = \pi_{\mathrm{Id}}(\sigma_{\mathrm{Type='Tank'}\wedge \mathrm{Faction='Enemy'}}(\mathit{R}))
\]

After merging the necessary partitions of relation $R$ and translating
it into positive relational algebra, we obtain 
\begin{align*}
\pi_{\mathrm{Id}}(\sigma_{\mathrm{Type='Tank'}\wedge \mathrm{Faction='Enemy'}}(U_1 \bowtie_{\alpha_1 \land \psi_1} U_2 \bowtie_{\alpha_2 \land \psi_2} U_3)),
\end{align*}
where the conditions $\psi_1$, $\psi_2$, $\alpha_1$, and $\alpha_2$
follow the translation given in Figure~\ref{fig:microops}. The three
vertical partitions are joined on the tuple id attributes ($\alpha_1$
and $\alpha_2$) and the combinations with conflicting mappings in the
ws-descriptors are discarded ($\psi_1$ and $\psi_2$).  Before and
after translation, the query is subject to optimizations as discussed
earlier. (In this case, a good query plan would first apply the
selections on the partitions, then project away the irrelevant value
attributes Type and Faction, and then merge the partitions).

\begin{center}{\small
    \bluebox{
    \begin{tabular}{@{}c@{ }|c@{ }c@{ }|c|cc}
      $U_4$ & $D_1$        & $D_2$        & T$_S$ & Id \\
      \hline
      & $x\mapsto 1$ &              & $c$ & 3\\
      & $x\mapsto 2$ &              & $c$ & 2\\
      & $y\mapsto 1$ & $z\mapsto 2$ & $d$ & 4
    \end{tabular}}}
\end{center}
The above U-relation $U_4$ encodes the query answer.\punto
\end{example}

\begin{example}
\label{ex:q-join}
\em We continue Example~\ref{ex:query} and ask whether it is possible
that the enemy has two tanks on the map, and if so, which vehicles are
those. For this, we compute the pairs of enemy tanks as a self-join of
$S$: $(S\ s_1) \bowtie_{s_1.\mathrm{Id}\not=s_2.\mathrm{Id}} (S\
s_2)$. This query is in turn equivalent to a self-join of $U_4$.
\begin{center}
  {\small
    \bluebox{
    \begin{tabular}{@{}c@{ }|c@{ }c@{ }c@{ }|c@{ }c|cc}
      $U_5$ & $D_1$        & $D_2$        & $D_3$        & T$_{s_1}$ & T$_{s_2}$ & Id$_1$ & Id$_2$ \\
      \hline
      & $x\mapsto 1$ & $y\mapsto 1$ & $z\mapsto 2$ & $c$       & $d$       & 3      & 4 \\
      & $x\mapsto 2$ & $y\mapsto 1$ & $z\mapsto 2$ & $c$       & $d$       & 2      & 4 \\
      & $y\mapsto 1$ & $z\mapsto 2$ & $x\mapsto 1$ & $d$       & $c$       & 4      & 3 \\
      & $y\mapsto 1$ & $z\mapsto 2$ & $x\mapsto 2$ & $d$       & $c$       & 4      & 2
    \end{tabular}}}
\end{center}
The answer is encoded by the above U-relation $U_5$. Note that the
combinations of the first two tuples of $U_4$ are not in $U_5$,
because they have inconsistent ws-descriptors and are filtered out
using the $\psi$-condition (vehicle $c$ cannot be at the same time at
two different positions). To obtain the possible pairs of vehicle ids,
we apply the poss operator on $U_5$. This is expressed as the
projection on the value attributes of $U_5$.\punto
\end{example}

Our translation yields relational algebra queries, whose evaluation always
produces tuple-level U-relations, i.e., U-relations without vertical
decompositions, by joining and merging vertical partitions of relations.
Following the definition of the merge operator, if the input U-relations are
reduced, then the result of merging vertical partitions is also reduced. We
thus have that

\begin{proposition}
  Given a positive relational algebra query $Q$ and a reduced
  U-relational database $U$, $\bananabr{Q}(U)$ is a reduced
  U-relational database.
\end{proposition}

%% file: normalization.tex
\section{Normalization of U-relations}
\label{sec:normalization}

\begin{algorithm}[t!]
{\small
        \KwIn{Reduced U-relational database ${\cal U}=(U_1, \dots, U_m, W)$}
        \KwOut{Normalized reduced U-relational database.}
        \Begin
        {
                $R$ := the
                relation consisting of all pairs of va\-ria\-bles
                $(c_i, c_j)$ that occur together in some ws-descriptor
                of ${\cal U}$\; 
                % or in two different ws-descriptors of the same U-relation which coincide on the tuple id vectors
                ${\cal G}$ := the graph whose node set is the set of
                variables and whose edge relation is
                the refl.\@ and trans.\@ closure of $R$\;
                Compute the connected components of ${\cal G}$\;% and assign ids to them\;
                \ForEach{U-relation $U_j(D_1, \ldots, D_n, \overline{T},
                \overline{A})$ of ${\cal U}$}
                {
                $U'_j$ := empty U-relation over
                $U'_j(\mbox{Var},\mbox{Rng},\overline{T},\overline{A})$\;
                \ForEach{$t \in U$}
                {
                        $G_i$ := connected component of ${\cal G}$
                with id $i$ such that
                the nodes $t.\mbox{Var}_1, \ldots, t.\mbox{Var}_n$ are in $G_i$\;
                        $\{c_{i_1}, \ldots, c_{i_k}\} = G_i - \{t.\mbox{Var}_1, \ldots, t.\mbox{Var}_n\}$\;
                        \ForEach{$l_{i_1}: (c_{i_1}, l_{i_1}) \in W,
                          \ldots, l_{i_k}: (c_{i_k}, l_{i_k}) \in W$}
                        {
                                /* Compute a new domain value ($f_{|G_i|}$
                is either
                the identity or better, for atomic $l$'s, an injective function
                $\mbox{int}^{|G_i|} \rightarrow \mbox{int}$) */\;
                                $l$ := $f_{|G_i|}(t.\overline{\mbox{Rng}},
                                         l_{i_1}, \ldots, l_{i_k})$\;
                                $U_j' := U_j' \cup \{ (G_i, l, t.\overline{T},
                                t.\overline{A}) \}$\;
                        }
                }
                }
        $W' := \bigcup_i \{ (g_i, (l_1, \dots, l_m)) \mid G_i = \{ c_1, \dots,
                c_m \}\mbox{ and}$\\
        \hspace{10.4em} $\; (c_1, l_1), \dots, (c_m, l_m) \in W \}$\;
        Output $(U_1', \dots, U_m', W')$\;
        }
}
\caption{Normalization of ws-descriptors.}
        \label{alg:norm}
\end{algorithm}

U-relations do not forbid large ws-descriptors. The ability to extend the size
of ws-descriptors is what yields efficient query evaluation on U-relations.
However, large ws-descriptors cause an inherent processing overhead. Also,
after query evaluation or dependency chasing on a U-relational data\-base, it
may happen that tuple fields, which used to be dependent on each other, become
independent. In such a case, it is desirable to optimize the world-set
representation~\cite{AKO07iWSD}. We next discuss one approach to normalize
U-relational databases by reducing large ws-descriptors to ws-descriptors of
size one. Normalization is an expensive operation per se, but it is not
unrealistic to assume that uncertain data is initially in normal form
\cite{AKO07WSD, AKO07iWSD} and can subsequently be maintained in this form.

\begin{definition}\label{def:wsdnf}
  \em A U-relational database is normalized if all ws-descriptors of its
  U-relations have size one.
\end{definition}

Algorithm~\ref{alg:norm} gives a normalization procedure for
U-relations that determines classes of variables that co-occur in some
ws-descriptors and replaces each such class by one variable, whose
domain becomes the product of the domains of the variables from that
class. Figure~\ref{fig:wsdnf-transformation} shows a U-relational
database and its normalization.

\begin{theorem}
  Given a reduced U-relational database, Algorithm~\ref{alg:norm}
  computes a normalized reduced U-relational database that represents
  the same world-set.
\end{theorem}

\nop{
WSDs and U-relations in WSDNF have the nice property that
a component can be put into a one-to-one correspondence with a set of
(tuple id, column name) pairs identifying fields in worlds
such that two tuple fields are
dependent within a component and independent across components.
This simplifies certain data cleaning problems (see \cite{AKO07WSD}).
} % end nop

\begin{figure}
\begin{small}
\begin{center}
\bluebox{
\begin{tabular}{c|@{\extracolsep{0.15cm}}c@{\extracolsep{0.15cm}}c|@{\extracolsep{0.15cm}}c|@{\extracolsep{0.15cm}}c}
$U$ & $D_1$ & $D_2$ & $T$ & $A$ \\
\hline 
& $c_1\mapsto 1$ & $c_1\mapsto 1$ & $t_1$ & $a_1$ \\
& $c_1\mapsto 1$ & $c_2\mapsto 2$ & $t_2$ & $a_2$ \\
& $c_1\mapsto 2$ & $c_1\mapsto 2$ & $t_2$ & $a_3$ \\
& $c_3\mapsto 1$ & $c_3\mapsto 1$ & $t_3$ & $a_4$ \\
& $c_3\mapsto 2$ & $c_3\mapsto 2$ & $t_3$ & $a_5$ \\
\end{tabular}
}
\hspace{1mm}
\bluebox{
\begin{tabular}{c|@{\extracolsep{0.15cm}}c@{\extracolsep{0.15cm}}c}
$W$ & Var & Rng \\
\hline
& $c_1$ & 1 \\
& $c_1$ & 2 \\
& $c_2$ & 1 \\
& $c_2$ & 2 \\
& $c_3$ & 1 \\
& $c_3$ & 2 \\
\end{tabular}
}
\smallskip

(a) U-relational database

\medskip
\bluebox{
\begin{tabular}{c|l|@{\extracolsep{0.15cm}}c|@{\extracolsep{0.15cm}}c@{\extracolsep{0.15cm}}}
$U'$ & $D$ & $T$ & $A$ \\
\hline 
& $c_{12}\mapsto (1,1)$ & $t_1$ & $a_1$ \\
& $c_{12}\mapsto (1,2)$ & $t_1$ & $a_1$ \\
& $c_{12}\mapsto (1,2)$ & $t_2$ & $a_2$ \\
& $c_{12}\mapsto (2,1)$ & $t_2$ & $a_3$ \\
& $c_{12}\mapsto (2,2)$ & $t_2$ & $a_3$ \\
& $c_3\mapsto 1$        & $t_3$ & $a_4$ \\
& $c_3\mapsto 2$        & $t_3$ & $a_5$ \\
\end{tabular}
}
\hspace{1mm}
\bluebox{
\begin{tabular}{c|@{\extracolsep{0.15cm}}l@{\extracolsep{0.15cm}}l@{\extracolsep{0.15cm}}}
$W'$ & Var & Rng \\
\hline
& $c_{12}$ & $(1,1)$ \\
& $c_{12}$ & $(1,2)$ \\ 
& $c_{12}$ & $(2,1)$ \\
& $c_{12}$ & $(2,2)$ \\ 
& $c_3$ & $1$ \\ 
& $c_3$ & $2$ \\ 
\end{tabular}
}
\smallskip

(b) Database from (a) normalized

\longversion{\medskip

\bluebox{
\begin{tabular}{|c|cc|}
\hline
$c_{12}$ & $t_1.A$ & $t_2.A$ \\
\hline
$(1,1)$ & $a_1$  & $\bot$ \\
$(1,2)$ & $a_1$  & $a_2$ \\
$(2,1)$ & $\bot$ & $a_3$ \\
$(2,2)$ & $\bot$ & $a_3$ \\
\hline
\end{tabular}
$\times$
\begin{tabular}{|c|c|}
\hline
$c_3$ & $t_3.A$ \\
\hline
1 & $a_4$ \\
2 & $a_5$ \\
\hline
\end{tabular}
}
\smallskip

(c) WSD corresponding to (b)}
\end{center}
\end{small}

\vspace{-1em}

\caption{Normalization example.}
\label{fig:wsdnf-transformation}
\vspace{-1em}
\end{figure}

{\noindent\bf Computing certain answers.}  Given a set of possible
worlds, we call a tuple certain iff it occurs in each of the worlds.
It is known that the tuple certainty problem is coNP-hard for a number
of representation systems, ranging from attribute-level ones like WSDs
to tuple-level ones like ULDBs~\cite{AKO07iWSD}. In case of
tuple-level normalized U-relations, however, we can efficiently
compute the certain tuples using relational algebra.

\begin{lemma}
  \em A tuple $\overline{t}$ is certain in a tuple-level normalized
  U-relation $U$ iff there exists a variable $x$ such that $(x\mapsto
  l,\overline{s},\overline{t})\in U$ for each domain value $l$ of $x$
  and some tuple id $\overline{s}$.
\end{lemma}

The condition of the lemma can be encoded as the following domain
calculus expression:
\begin{align*}
cert(U) := \{\overline{t}\ |\ \exists x \forall l\; (x,l) \in W \Rightarrow \exists \overline{s} (x,l,\overline{s},\overline{t}) \in U \}
\end{align*}
The equivalent relational algebra query on a tuple-level normalized
U-relational database
$(U[\mbox{Var},\mbox{Rng},\overline{T_R},\overline{A}],W)$ is
\begin{align*}
   \pi_{\overline{A}}(\pi_{\mathit{Var}}(W) \times \pi_{\overline A}(U) - \pi_{\mathit{Var}, \overline{A}}(W \times \pi_{\overline{A}}(U) - \pi_{\mathit{Var},\mathit{Rng},\overline{A}}U)).
\end{align*}

\vspace{-2mm}

%%% Local Variables: 
%%% mode: latex
%%% TeX-master: "icde2008_maybms"
%%% End: 

%% file: discussion.tex
\section{Succinctness and Efficiency}
%\section{Discussion: UWSDs, ULDBs, and U-Relations}
%\section{UWSDs, ULDBs, and U-Relations}
\label{sec:discussion}

This section compares U-relational databases with WSDs
\cite{AKO07WSD,AKO07iWSD} and ULDBs \cite{BDSHW2006} using two
yardsticks: succinctness, i.e., how compactly can they represent
world-sets, and efficiency of query evaluation.  \shortversion{Due to
  lack of space, we defer a more complete comparison (with proofs and
  examples) to an extended version of this paper~\cite{AKO07-Urel}.}

\smallskip{\noindent\bf WSDs vs.\ U-Relations.} WSDs are essentially normalized
U-relational databases where each variable $c_i$ of a U-relation
corresponds to a WSD \textit{component} relation $C_i$ and each domain
value $l_i$ of $c_i$ corresponds to a tuple of
$C_i$. \longversion{Figure~\ref{fig:wsdnf-transformation}(c) shows a
WSD equivalent to a normalized U-relational database.} The
normalization may lead to an exponential blow-up in the database size
and accounts for U-relations with arbitrarily large ws-descriptors
being more compact than U-relations with singleton ws-descriptors and
thus than WSDs.

\longversion{
\begin{figure}[tbp]
  \centering{\small
    \subfigure[WSD encoding.]{
\bluebox{
    \begin{tabular}{|l|cc|}\hline
      $c_1$ & $t_1.A$ & $t_2.B$\\\hline
      $w_1$ & $1$     & $1$\\
      $w_2$ & $0$     & $0$\\\hline
    \end{tabular}
%  $\times$
%  \begin{tabular}{|l|ll|}\hline
%    $c_2$ & $t_2.A$ & $t_3.B$\\\hline
%    $w_1$ & $1$     & $1$\\\hline
%    $w_2$ & $0$     & $0$\\\hline
%  \end{tabular}%
    $\times\cdots\times$
    \begin{tabular}{|l|cc|}\hline
      $c_n$ & $t_n.A$ & $t_1.B$\\\hline
      $w_1$ & $1$     & $1$\\
      $w_2$ & $0$     & $0$\\\hline
    \end{tabular}}}

  \subfigure[U-relational encoding.]{\hspace*{-1em}%
\bluebox{
  \begin{tabular}{l|c|l|l}
    $U_1$ & $D$                & $T$   & $A$\\\hline
          & $c_1\mapsto w_1$   & $t_1$ & $1$\\
          & $c_1\mapsto w_2$   & $t_1$ & $0$\\
          & $\vdots$           &       &     \\
          & $c_n\mapsto w_1$   & $t_n$ & $1$\\
          & $c_n\mapsto w_2$   & $t_n$ & $0$\\
  \end{tabular}}%
  \hspace*{1em}%
  \bluebox{
  \begin{tabular}{l|c|l|l}
    $U_2$ & $D$                & $T$   & $B$\\\hline
          & $c_2\mapsto w_1$   & $t_2$ & $1$\\
          & $c_2\mapsto w_2$   & $t_2$ & $0$\\
          & $\vdots$           &       & \\
          & $c_n\mapsto w_1$   & $t_1$ & $1$\\
          & $c_n\mapsto w_2$   & $t_1$ & $0$\\
  \end{tabular}}}}

  \vspace*{-1em}

  \caption{WSD and U-relational encoding of the world-set of Example~\ref{ex:wsd-urel}.}
  \label{fig:wsd-urel}
  \vspace*{-1em}
\end{figure}

\begin{example}\em\label{ex:wsd-urel}
  Consider a relation over schema $R[AB]$ where each field value can be 0 or
  1, and $t_i.A$ and the tuple fields $t_{(i+1) \mbox{ mod } n}.B$ depend on
  each other ($1\leq i\leq n$). The encodings as WSD and as a set of two
  U-relations are given in Figure~\ref{fig:wsd-urel}.\punto
\end{example}}

\begin{theorem}\label{th:wsd-succinctness}
  U-relational databases are exponentially more succinct than WSDs.
\end{theorem}

Positive relational queries have polynomial data complexity for
U-relations (Section~\ref{sec:queries}) and exponential data
complexity for WSDs~\cite{AKO07iWSD}. This can be explained in close
analogy to the difference in succinctness and by the fact that query
evaluation creates new dependencies \cite{dalvi04efficient}:
U-relations can efficiently store the new dependencies by enlarging
ws-descriptors, whereas WSDs correspond to U-relations with normalized
ws-descriptors, hence the exponential blowup.

\longversion{
\begin{figure}[tbp]
  \centering{\small
    \subfigure[WSD encoding.]{
\bluebox{
    \begin{tabular}{|c|ccccc|}\hline
      $c_1\times\cdots\times c_n$ & $t_1.A$ & $t_2.B$ & $\ldots$ & $t_n.A$ & $t_1.B$\\\hline
      $w_1$                       & $1$     & $1$     & $\ldots$ & $1$     &  $1$   \\
      $w_2$                       & $0$     & $0$     & $\ldots$ & $0$     &  $0$   \\
      $w_3$                       & $\bot$  & $1$     & $\ldots$ & $1$     &  $\bot$   \\
                                  &         &         & $\vdots$ &         &        \\
      $w_{2^n}$               & $\bot$  & $\bot$  & $\ldots$ & $\bot$  & $\bot$ \\\hline
    \end{tabular}}}

  \subfigure[U-relational encoding.]{\hspace*{-1em}%
\bluebox{
  \begin{tabular}{c|cc|c|cc}
    $U_3$ & $D_1$             & $D_2$             & $T$   & $A$ & $B$\\\hline
          & $c_2\mapsto w_1$  & $c_3\mapsto w_1$  & $t_2$ & $1$ & $1$\\
          & $c_2\mapsto w_2$  & $c_3\mapsto w_2$  & $t_2$ & $0$ & $0$\\
          &                   & $\vdots$          &       &     &  \\
          & $c_1\mapsto w_1$  & $c_n\mapsto w_1$  & $t_1$ & $1$ & $1$\\
          & $c_1\mapsto w_2$  & $c_n\mapsto w_2$  & $t_1$ & $0$ & $0$\\
  \end{tabular}}}}

  \vspace*{-1em}

  \caption{WSD and U-relation representing the answer to $\sigma_{A=B}(R)$ with $R$ of Figure~\ref{fig:wsd-urel}.}
  \label{fig:wsd-urel2}
  \vspace*{-1em}
\end{figure}

\begin{example}\em\label{ex:wsd-blowup}
  Consider the WSD and U-relations of Example~\ref{ex:wsd-urel} and
  the selection with join condition $\sigma_{A=B}(R)$. The answer is
  represented by the WSD and U-relation respectively shown in
  Figure~\ref{fig:wsd-urel2}.  The U-relation $U_3$ has $2\cdot n$
  tuples, whereas the WSD $c_1\times\cdots\times c_n$ has $2^n$
  tuples, each representing a possible combination of the values of
  the existing fields (a tuple $t_i$ does not occur in worlds where
  $t_i.A$ or $t_i.B$ have values $\bot$). Note that by normalizing
  $U_3$ we would also obtain one variable with $2^n$ domain values, as
  for the WSD.
  
  The answer to $\mathit{poss}(\sigma_{A=B}(R))$ is efficiently
  computed as $\pi_{A,B}(U_3)$ in the case of U-relations.
  In the WSD case, it is computed as
  $\underset{i}{\overset{n}{\cup}}(\pi_{t_i.A,t_i.B}(c_1\times\cdots\times
  c_n))$.  \punto
\end{example}

Finally, the query translations employed by the evaluation algorithms
in the WSD and U-relational cases are different. Whereas for WSDs all
operators are translated to sequences of relational queries and in the
case of projection and join even to fixpoint programs~\cite{AKO07WSD},
the translation remains strictly in relational algebra for
U-relations.}

%\begin{definition}
%A UWSD is essentially a U-relation in WSDNF.
%\end{definition}

%Make this precise.
%Note that UWSDs were defined in \cite{AKO07WSD} directly; so this definition
%is really a theorem that, however, holds by construction.

%Using template relations does not pay off anymore when
%we process queries using the methods of this paper, so we store the entire
%possible-worlds database using U-relations. Explain why.

%Discuss the conceptual simplicity of representation and querying using
%U-relations. Queries involving selection,
%projection, join and possible on U-relations can be
%encoded in relational algebra in a simple way while query processing
%on WSDs sometimes seems quite complicated.

%\begin{example}\em
%Show how, for a small WSD, projection seems quite complicated but becomes
%very intuitive when executed as a semijoin on the U-relation representation.
%\end{example}

%Discuss why selection and projection on WSDs in general
%require exponential blowup while this does not happen for U-relations.
%Explain why the transformation of U-relations into WSDNF is exponential but
%why we do not need to perform it to evaluate positive queries.

%Discuss the importance of attribute-level representations of uncertainty and
%that UWSDs and thus U-relations support them (give an example!)
%but that ULDBs don't.

\smallskip{\noindent\bf ULDBs vs.\ U-Relations.} 
\shortversion{ULDBs are databases with
  uncertainty and lineage~\cite{BDSHW2006}. Due to lack of space, we
  only state the salient results concerning our comparison to ULDBs.}
\longversion{
A ULDB relation is a set of \textit{x-tuples}, where each x-tuple
represents a set of alternatives. One world is defined by choosing
precisely one alternative of each x-tuple. A world may contain none of
the alternatives of an x-tuple, if this x-tuple is marked as optional
(or \textit{maybe}) using the \textbf{?}-symbol.  Dependencies between
alternatives of different x-tuples are enforced using
\textit{lineage}: An alternative $i$ of an x-tuple $s$ occurs in the
same worlds with an alternative $j$ of another x-tuple $t$ if the
lineage of $(s,i)$ points either to $(t,j)$, or to another alternative
that transitively points to $(t,j)$. The lineage of an alternative can
also point to an \textit{external} symbol $(t,j)$, if there is no
alternative $(t,j)$ in the database~\cite{BDSHW2006}.

\begin{example}\em\label{ex:uldb1}
  The U-relations representing relation $R$ in Figure~\ref{fig:u-rel} admit
  the following equivalent ULDB:

\input{trio-example}
\vspace*{-1em}

To construct an ULDB equivalent to the U-relational database of
Figure~\ref{fig:u-rel}, we have to enumerate all possible value
combinations for the attributes of $R$. This enumeration is not
necessary for U-relations because of vertical partitioning and the
independence of (most) tuple fields. \punto
\end{example}}

\begin{lemma}
  ULDBs~\cite{BDSHW2006} can be translated linearly into U-relational
  databases.  \longversion{\begin{proof} We sketch the proof for a
      single ULDB relation $R$; it can be extended trivially to the
      case of several relations.
    
    For every x-tuple $t$ in $R$ we create a new variable $c_t$, and for each
    alternative $j$ of $t$ we create a new domain value $w_{(t,j)}$ of $c_t$.
    For every alternative in $R$ with value $a$, id $(t,j)$ and lineage
    $\lambda(t,j)=\overset{n_{(t,j)}}{\underset{i}{\bigwedge}}(t_i,j_i)$ we
    create a tuple in $U^R$ with value $a$, tuple id $t$ and ws-descriptor
    ($n = n_{(t,j)}$)
    $$D_{(t,j)} =
    [(c_t,w_{(t,j)}),(c_{t_1},w_{(t_1,j_1)}),\ldots,(c_{t_n},w_{(t_n,j_n)})].$$
    In case $n_{(t,j)}$ is smaller than $n_{(s,l)}$ of an alternative $l$ of
    an x-tuple $s$, then we pad the above ws-descriptor with
    $n_{(s,l)}-n_{(t,j)}$ pairs $(c_t,w_{(t,j)})$.
    
    The world table $W$ is the set of pairs of variables and domain values
    created for the x-tuples of $R$. For each optional x-tuple $t$ in $R$, we
    also add to $W$ a tuple $(c_t,w)$ where $w$ is a fresh domain value for
    $c_t$.
  \end{proof}}
\end{lemma}

\shortversion{The translation uses a direct encoding of ULDB's lineage
  into ws-descriptors, where ULDB's tuple and alternative ids become
  variables and domain values, respectively.}

%% The generic translation of ULDBs to U-relations used in the proof of
%% Theorem~\ref{th:uldb-succinctness} allows existing data
%% cleaning algorithms for WSDs \cite{AKO07WSD} (and thus U-relations in WSDNF)
%% to be carried over to ULDBs. Similarly, the computation of certain
%% answers (cf.\@ Section~\ref{sec:certain}) can be used for ULDBs. This
%% is interesting as, to date, there are no such algorithms for
%% ULDBs.

There are U-relations, however, whose ULDB encodings are necessarily
exponential in the arity of the logical relation. This is the case of,
e.g., or-set relations~\cite{INV1991}, attribute-level representations
that can be linearly encoded as U-relations but exponentially as
ULDBs.

\begin{theorem}\label{th:uldb-succinctness}
  U-relational databases are exponentially more succinct than ULDBs.
\end{theorem}

% Or-set relations are attribute-level and thus represented by
%    exponentially larger ULDBs and by WSDs of linear size~\cite{AKO07WSD}.

Both ULDBs and U-relations have polynomial data complexity for
positive relational queries. Differently from ULDBs, evaluating
queries on U-relations is possible using relational algebra only.  The
main difference between their evaluation algorithms concerns erroneous
tuples, i.e., tuples that do not appear in any world. In contrast to
U-relations, erroneous tuples may appear in the answers to queries on
ULDBs (see \cite{BDSHW2006} for an example). The removal of such
tuples is called data minimization, an expensive operation that
involves the computation of the transitive closure of
lineage~\cite{BDSHW2006}. Such tuples occur with ULDBs because the
lineage of an alternative in the answer only points to the lineage of
alternatives from the input relations, even though these input
alternatives may not occur in the same world. This cannot happen with
U-relations because each query operation ensures that only valid
tuples are in the query answer by (1) using the $\psi$-condition in
the join and merge operations and by (2) carrying all dependencies in
the ws-descriptors -- and not only to tuples of the input relation.

%\begin{remark}\em
%  ULDBs are a syntactically restricted form of c-tables~\cite{IL1984}, where
%  x-tuple alternatives are c-tuples without variables and the lineage of
%  alternatives is encoded as local boolean conditions. The mutual
%  exclusiveness of alternatives of an x-tuple is captured by additional local
%  conditions such that precisely one of these conditions is true under any
%  valuation of the variables in the conditions.\punto
%\end{remark}

\longversion{\medskip

  To sum up, U-relations have the advantages of WSDs (attribute-level
  representation) and ULDBs (polynomial evaluation of positive
  relational algebra queries), while forming an exponentially more
  succinct representation system than both aforementioned approaches.}

%% file: trio-example.tex
  \begin{center}
    \hspace*{-1em}{\small\bluebox{
    \begin{tabular}{l|ll|l}
          & \hspace*{1em} R (Id, Type, Faction) & & \\\hline
    $a$ & 1: (1, Tank, Friend) &  \\\hline
    $b$ & 1: (2, Transport, Friend) & $||$ 2: (3, Transport, Friend) & $\Lambda$ \\\hline
    $c$ & 1: (3, Tank, Enemy)       & $||$ 2: (2, Tank, Enemy) & \\\hline
    $d$ & 1: (4, Tank, Friend)      & $||$ 2: (4, Tank, Enemy) $||$ & \\
        & 3: (4, Transport, Friend) & $||$ 4: (4, Transport, Enemy) & \\\hline
    \end{tabular}}\\
    \begin{align*}
      \Lambda \mbox{ is } & \lambda(b,1)=\{(c,1)\},\lambda(b,2)=\{(c,2)\}\\
    \end{align*}}
\end{center}

%%% Local Variables: 
%%% mode: latex
%%% TeX-master: "icde2008_maybms"
%%% End: 

%% file: experiments.tex
\begin{figure}[t!]
        \begin{center}
          \framebox{ \parbox{8cm}{\footnotesize $Q_1$: \textbf{possible} (\textbf{select}
              o.orderkey, o.orderdate, o.shippriority \textbf{from}
              customer c, orders o, lineitem l \textbf{where}              c.mktsegment $=$ 'BUILDING'\\
              \textbf{and} c.custkey $=$ o.custkey and o.orderkey $=$ l.orderkey\\
              \textbf{and} o.orderdate $>$ '1995-03-15' \textbf{and} l.shipdate $<$ '1995-03-17')\\
 
                $Q_2$: \textbf{possible} (\textbf{select} extendedprice \textbf{from} lineitem \textbf{where}\\
                shipdate \textbf{between} '1994-01-01' \textbf{and} '1996-01-01'\\
                \textbf{and} discount \textbf{between} '0.05' \textbf{and} '0.08' \textbf{and} quantity $<$ 24)\\
               
                $Q_3$: \textbf{possible} (\textbf{select} n1.name, n2.name \textbf{from}
                supplier s, lineitem l,\\
                orders o, customer c, nation n1, nation n2 \textbf{where} n2.nation='IRAQ'\\
                \textbf{and} n1.nation='GERMANY' \textbf{and} c.nationkey $=$ n2.nationkey\\
                \textbf{and} s.suppkey $=$ l.suppkey \textbf{and} o.orderkey $=$ l.orderkey\\
                \textbf{and} c.custkey $=$ o.custkey \textbf{and} s.nationkey $=$ n1.nationkey)
        }
        }
    \end{center}
  \vspace*{-1em}

  \caption{Queries used in the experiments.}
  \label{fig:exp-query}
  \vspace*{-1.5em}
\end{figure}

\shortversion{
\begin{figure*}[t!]
  \begin{footnotesize}
\begin{tabular}{|c|c||@{~}r|@{\extracolsep{0.2cm}}l@{\extracolsep{0.2cm}}r@{\extracolsep{0.2cm}}r|@{\extracolsep{0.2cm}}l@{\extracolsep{0.2cm}}r@{\extracolsep{0.2cm}}r|@{\extracolsep{0.2cm}}l@{\extracolsep{0.2cm}}r@{\extracolsep{0.2cm}}r|}\hline
              &   & TPC-H  &          &       &        &        &     &        &           &     &\\
            s & z & dbsize & \#worlds & Rng & dbsize & \#worlds & Rng & dbsize &  \#worlds & Rng & dbsize\\\hline
                        0.01 & 0.1 & 17  & $10^{857.076}$ & 21 & 82 & $10^{7955.30}$ & 57 & 85  & $10^{79354.1}$ & 57 & 114 \\
                        0.01 & 0.5 & 17 & $10^{523.031}$ & 71 & 82 & $10^{4724.56}$ & 901 & 88 & $10^{46675.6}$ & 662 & 139 \\\hline
            0.05 & 0.1 & 85 & $10^{4287.23}$ & 22 & 389 & $10^{39913.8}$ & 33 & 403 & $10^{396137}$ & 65 & 547 \\
                        0.05 & 0.5 & 85 & $10^{2549.14}$ & 178& 390 & $10^{23515.5}$ & 449 & 416 & $10^{232650}$ & 1155 & 672 \\\hline
            0.10 & 0.1 & 170 & $10^{8606.77}$ & 27 & 773& $10^{79889.9}$ & 49 & 802 & $10^{793611}$ & 53 & 1090 \\
                        0.10 & 0.5 & 170 & $10^{5044.65}$ & 181 & 776 & $10^{46901.8}$ & 773 & 826 & $10^{466038}$ & 924 & 1339 \\\hline
                        0.50 & 0.1 & 853 & $10^{43368.0}$ & 49 & 3843 & $10^{400185}$ & 71 & 3987 & $10^{3.97e+06}$ & 85 & 5427 \\
                        0.50 & 0.5 & 853 & $10^{25528.9}$ & 214 & 3856 & $10^{234840}$ & 1832 & 4012 & $10^{2.33e+06}$ & 2586 & 6682 
                        \\\hline
            1.00 & 0.1 & 1706 & $10^{87203.0}$ & 57 & 7683 & $10^{800997}$ & 99 & 7971 & $10^{7.94e+06}$ & 113 & 11264 \\
                        1.00 & 0.5 & 1706 & $10^{51290.9}$ & 993 & 7712 & $10^{470401}$ & 1675 & 8228 & $10^{4.66e+06}$ & 3392 & 13312 \\\hline\hline
                        & & $\mathbf{x} = 0.0$ & \multicolumn{3}{c|}{$\mathbf{x} = 0.001$} & \multicolumn{3}{c|}{$\mathbf{x} = 0.01$} &  \multicolumn{3}{c|}{$\mathbf{x} = 0.1$}\\\hline
\end{tabular}

\vspace{-4.8cm}

{ }~{ }\hspace{12.2cm}
\input{bigtree}

\medskip

\end{footnotesize}
  \vspace*{-1em}
  \caption{(left): Total number of worlds, max.\ number of domain values for a
    variable (Rng), and size in MB of the U-relational database for each of
    our settings. (right): Query plan for $Q_1$ using merge.}
  \label{fig:dbsize}
 \label{fig:translevel2}
  \vspace*{-1.5em}
\end{figure*}
}

\longversion{
\begin{figure*}[t!]
  \begin{small}
  \begin{center}
  \begin{tabular}{|l|l||r|lrr|lrr|lrr|}\hline
     scale & correlation  & TPC-H dbsize & \#worlds & lworlds & dbsize & \#worlds & lworlds & dbsize &  \#worlds & lworlds & dbsize\\\hline
     0.01 & 0.10 & 17   & $10^{857.076}$   & 21   & 82   & $10^{7955.30}$   & 57    & 85   & $10^{79354.1}$   & 57    & 114 \\
     0.01 & 0.25 & 17   & $10^{729.529}$   & 33   & 82   & $10^{6728.24}$   & 129   & 85   & $10^{66995.5}$   & 193   & 118 \\
     0.01 & 0.50 & 17   & $10^{523.031}$   & 71   & 82   & $10^{4724.56}$   & 901   & 88   & $10^{46675.6}$   & 662   & 139 \\\hline
     0.05 & 0.10 & 85   & $10^{4287.23}$   & 22   & 389  & $10^{39913.8}$   & 33    & 403  & $10^{396137}$    & 65    & 547 \\
     0.05 & 0.25 & 85   & $10^{3633.49}$   & 57   & 389  & $10^{33702.3}$   & 148   & 405  & $10^{334450}$    & 158   & 567 \\
     0.05 & 0.50 & 85   & $10^{2549.14}$   & 178  & 390  & $10^{23515.5}$   & 449   & 416  & $10^{232650}$    & 1155  & 672 \\\hline
     0.10 & 0.10 & 170  & $10^{8606.77}$   & 27   & 773  & $10^{79889.9}$   & 49    & 802  & $10^{793611}$    & 53    & 1090 \\
     0.10 & 0.25 & 170  & $10^{7276.46}$   & 74   & 774  & $10^{67477.1}$   & 145   & 806  & $10^{670090}$    & 172   & 1132 \\
     0.10 & 0.50 & 170  & $10^{5044.65}$   & 181  & 776  & $10^{46901.8}$   & 773   & 826  & $10^{466038}$    & 924   & 1339 \\\hline
     %     &      &      &                  &      &      &                  &       &      &                  &       &      \\
     0.50 & 0.10 & 853  & $10^{43368.0}$   & 49   & 3843 & $10^{400185}$   & 71   & 3987 & $10^{3.96845e+06}$   & 85   & 5427 \\
     0.50 & 0.25 & 853  & $10^{36630.3}$   & 130  & 3845 & $10^{337905}$   & 172  & 4008 & $10^{3.35095e+06}$   & 320  & 5632 \\
     0.50 & 0.50 & 853  & $10^{25528.9}$   & 214  & 3856 & $10^{234840}$   & 1832 & 4012 & $10^{2.33083e+06}$   & 2586 & 6682 \\\hline
     1.00 & 0.10 & 1706 & $10^{87203.0}$   & 57   & 7683 & $10^{800997}$   & 99   & 7971 & $10^{7.93774e+06}$   & 113  & 11264 \\
     1.00 & 0.25 & 1706 & $10^{73652.5}$   & 170  & 7687 & $10^{676223}$   & 208  & 8012 & $10^{6.70229e+06}$   & 344  & 11280 \\
     1.00 & 0.50 & 1706 & $10^{51290.9}$   & 993  & 7712 & $10^{470401}$   & 1675 & 8228 & $10^{4.66222e+06}$   & 3392 & 13312 \\\hline\hline
     &      & $\mathbf{x} = 0.0$ & \multicolumn{3}{c|}{$\mathbf{x} = 0.001$} & \multicolumn{3}{c|}{$\mathbf{x} = 0.01$} &  \multicolumn{3}{c|}{$\mathbf{x} = 0.1$}\\\hline
   \end{tabular}
 \end{center}
        \end{small}
        \vspace*{-1.5em}
  \caption{Total number of worlds, max.\ number of local worlds in a component, and size in MB of the U-relational database for each of our settings.}
  \label{fig:dbsize}
  \vspace*{-1em}
\end{figure*}
}

\section{Experiments}
\label{sec:experiments}

\noindent{\bf Prototype Implementation.} We implemented the query
translator of Figure~\ref{fig:microops} and also extended the C
implementation of the TPC-H population generator version 2.6 build
1~\cite{tpch2006} to generate attribute and tuple-level U-relations
and ULDBs. The code is available on the MayBMS project page
(\textrm{http://www.infosys.uni-sb.de/projects/maybms}).

\noindent{\bf Setup.} The experiments were performed on a 3GHZ/1GB
Pentium running Linux 2.6.13 and PostgreSQL 8.2.3.

\noindent{\bf Generation of uncertain data.}
\longversion{ Our data generator creates eight tables: part, partsupp,
  supplier, customer, lineitem, orders, nation, region. The field
  values are sensitive to the attribute types and are randomly
  generated or randomly chosen from the dictionary explained in the
  TPC-H benchmark specification.  }
The following parameters were used to tune the generation: {\it
  scale\/} ($s$), {\it uncertainty ratio\/} ($x$), {\it correlation
  ratio\/} ($z$), and {\it maximum alternatives per field\/} ($m$).
The (dbgen standard) parameter $s$ is used to control the size of each
world; $x$ controls the percentage of (uncertain) fields with several
possible values, and $m$ controls how many possible values can be
assigned to a field. The parameter $z$ defines a Zipf distribution for
the variables with different dependent field counts\footnote{This is
  the number of tuple fields dependent on that variable.} (DFC) and
controls the attribute correlations: For $n$ uncertain fields, there
are $\lceil C*z^{i} \rceil$ variables with DFC $i$, where $C =
n(z-1)/(z^{k+1} - 1)$, i.e., $n = \overset{k}{\underset{i=0}{\Sigma}}
(C*z^i)$.  The number of domain values of a variable with DFC $k>1$ is
chosen using the formula
$p^{k-1}*\overset{k}{\underset{i=1}{\Pi}}(m_i)$, where $m_i$ is the
number of different values for the field $i$ dependent on that
variable and $p$ is the probability that a combination of possible
values for the $k$ fields occurs. This assumption fits naturally to
data cleaning scenarios. Previous work~\cite{AKO07WSD} shows that
chasing dependencies on WSDs enforces correlations between field
values and removes combinations that violate the dependencies.  We
considered here that after correlating two variables with arbitrary
DFCs, $100(1-p)$ percent of the combinations violate constraints and
thus are dropped. 

The uncertain fields are assigned randomly to variables. This can lead
to correlations between fields belonging to different tuples or even
to different relations. This fits to scenarios where constraints are
enforced across tuples or relations. We do not assume any kind of
independence of our initial data as done in several other
approaches~\cite{dalvi04efficient,BDSHW2006}.

\longversion{
Our data generator works as follows.  While generating tuples for the
eight tables, we use the uncertainty ratio to decide at each tuple
field if it is uncertain or not. We collect in a field pool the
coordinates (i.e., relation, tuple id, attribute) of the uncertain
tuple fields and when the original TPC-H generator finishes its job or
the field pool is full, we shuffle the uncertain tuple fields, compute
the correlation ratio for variables with different DFC, and
incrementally assign tuple fields to variables. Then, we compute the
domain size of each variable, and the number of different values for
each of variable's fields. The field values are then generated using
the data distribution and dictionary for that field type, as specified
by the original TPC-H generator. Because there can be too many field
coordinates to keep in memory at a time, we use in our experiments a
window of 10 million fields to be processed in bulk\footnote{It
corresponds to a maximum of 500 MB of main memory allocated for dbgen
on our testing machine.}; after a window is processed, the memory is
released, and a new window is filled in and processed.  The window
size influences the number and dependent field count of the variables.
}
For the experiments, we fixed $p$ to 0.25, $m$ to 8, and varied the
remaining parameters as follows: $s$ ranges over $(0.01, 0.05, 0.1,
0.5, 1)$, $z$ ranges over $(0.1, 0.25, 0.5)$, and $x$ ranges over
$(0.001, 0.01, 0.1)$.

An important property of our generator is that any world in a
U-relational database shares the properties of the one-world database
generated by the original dbgen: The sizes of relations are the same
and the join selectivities are approximately equal. We checked this by
randomly choosing one world of the U-relational database and comparing
the selectivities of joins on the keys of the TPC-H relations for
different scale factors and uncertainty ratios.

\longversion{
 \begin{figure}[t!]
   \centering
\input{bigtree}
 \caption{Query plan for $Q_1$ using merge.}
 \label{fig:translevel2}
 \vspace{-0.6cm}
 \end{figure}
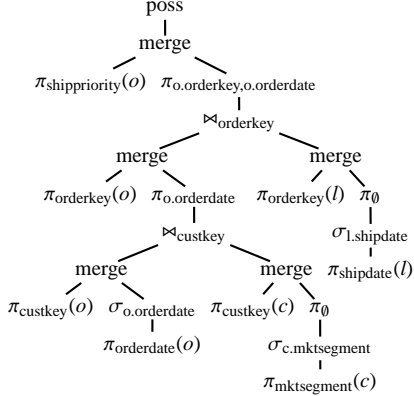}

\begin{figure*}[t!]
        \begin{center}
        \begin{tabular}{c@{\ }cc}
     \includegraphics[scale=1]{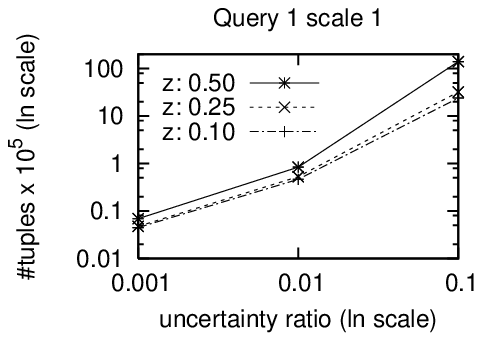}
     &
     \includegraphics[scale=1]{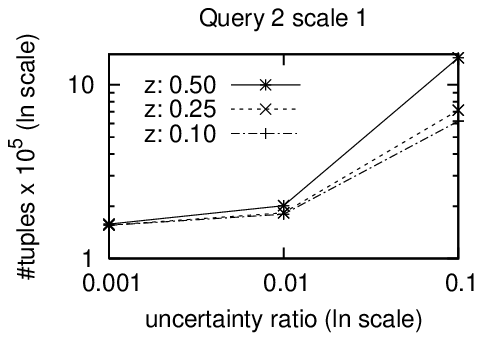}
     &
     \includegraphics[scale=1]{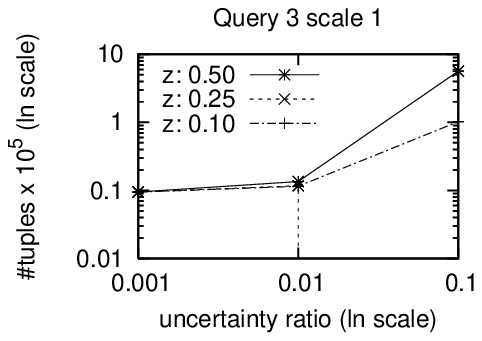}
  \end{tabular}
  \end{center}
  \vspace*{-8mm}

  \caption{Sizes of query answers for settings with scale 1.}
  \label{fig:answersizes}
\end{figure*}

%\medskip

\noindent{\bf Queries.} We used the three queries from
Figure~\ref{fig:exp-query}. Query $Q_1$ is a join of three relations
of large sizes.  Query $Q_2$ is a select-project query on the relation
lineitem (the largest in our settings). Query $Q_3$ is a fairly
complex query that involves joins between six relations. All queries
use the operator `possible' to retrieve the set of matches across all
worlds.  Note that these queries are modified versions of $Q_3$,
$Q_6$, and $Q_7$ of TPC-H where all aggregations are dropped (dealing
with aggregation is subject to future work).

Figure~\ref{fig:answersizes} shows that our queries are moderately
selective and their answer sizes increase with uncertainty $x$ and
marginally with correlation $z$. For scale 1, the answer sizes range
from tens of thousands to tens of millions of tuples.  There is only one
setting ($z = 0.25$ and $x = 0.1$) where one of our queries, $Q_3$,
has an empty answer.  Before the execution, the queries were optimized
using our U-relation-aware optimizations.  Figure
\ref{fig:translevel2} shows $Q_1$ after optimizations.

\shortversion{
\begin{figure*}[t!]
  \begin{center}
    \vspace{-1em}
        \begin{tabular}{c@{\ }cc}
        \includegraphics[scale=1]{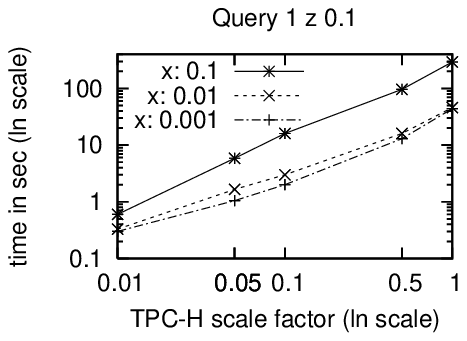}
        &
        \includegraphics[scale=1]{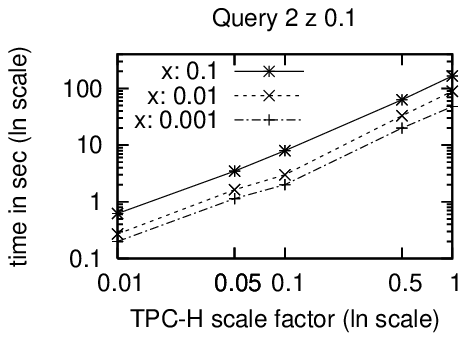}
        &
        \includegraphics[scale=1]{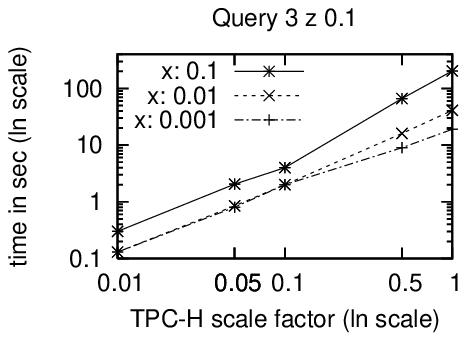}
        \\
        \includegraphics[scale=1]{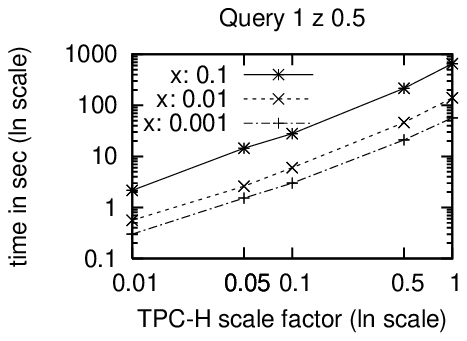}
        &
        \includegraphics[scale=1]{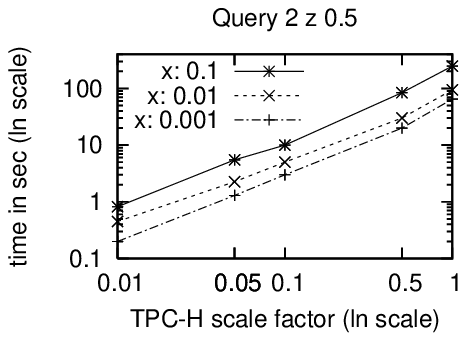}
        &
        \includegraphics[scale=1]{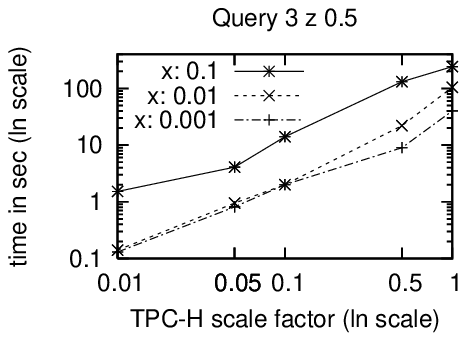}
        \end{tabular}
  \end{center}
 \vspace*{-6mm}
 \caption{Performance of query evaluation for various scale, uncertainty, and correlation.}
 \label{fig:perf}
 \vspace*{-1.5em}
\end{figure*}
}

\longversion{
        \begin{figure*}[t!]
    \begin{center}
    \begin{tabular}{c@{}cc}
           \epsfig{file=fig/query1z0_1.eps}
           &
           \epsfig{file=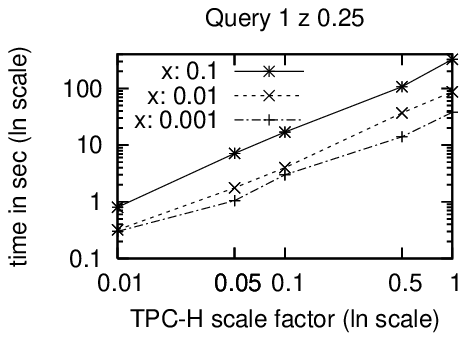}
           &
           \epsfig{file=fig/query1z0_5.eps}\\
           \epsfig{file=fig/query2z0_1.eps}
           &
           \epsfig{file=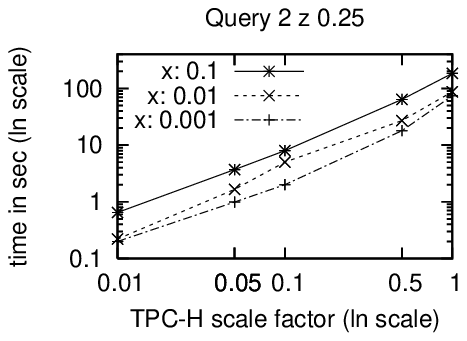}
           &
           \epsfig{file=fig/query2z0_5.eps}\\
           \epsfig{file=fig/query3z0_1.eps}
           &
            \epsfig{file=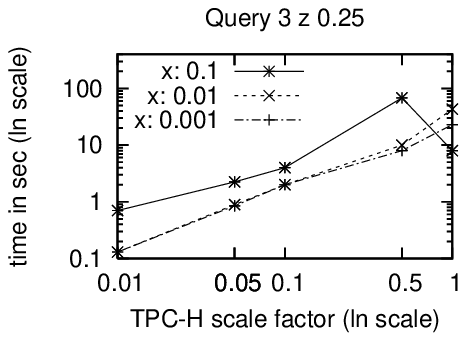}
           &
           \epsfig{file=fig/query3z0_5.eps}
        \end{tabular}
        \end{center}
        \vspace*{-1em}
        
          \caption{Performance of query evaluation for various scale, uncertainty, and correlation.}
          \label{fig:perf}
        \vspace*{-1.5em}
        \end{figure*}
}

%%%%%%%%%%%%%%%%%%%%%%%%%%%%%%%%%%%%%%%

\noindent{\bf Characteristics of U-relations.} Following Figure~\ref{fig:dbsize}, the U-relational
databases are exponentially more succinct than databases representing
all worlds individually: while the number of worlds increases
exponentially (when varying the uncertainty ratio $x$), the database
size increases only linearly. The case of $x=0$ corresponds to one
world generated using the original dbgen. Interestingly, to represent
$10^{8\cdot 10^6}$ worlds, the U-relational database needs about 6.7
times the size of one world.

An increase of the scaling factor leads to an exponential increase in
the number of worlds and only to a linear increase in the size of the
U-relational database. \longversion{The maximum domain size of a
variable is indirectly influenced by $s$: When $s$ increases, there
are more uncertain fields and thus more likely to obtain variables
with more dependent fields. By our construction, the domain size of
variables with higher DFC can be much larger than the maximum domain
size of variables with DFC=1 (which is $m=8$).  This is because a
variable with DFC=$k$ has a fraction ($p=0.25)$ of the product of the
domain values of $k$ variables taken together. As shown in
Figure~\ref{fig:dbsize}, our settings have variables with domain sizes
of up to 3392.} Although we only report here on experiments with scale
factors up to 1, further experiments confirmed that similar
characteristics are obtained for larger scales, too. An increase of
the correlation parameter leads to a moderate relative increase in the
database size. When compared to one-world databases, the sizes of
U-relational databases have increase factors that vary from 6.2 (for
$z=0.1$) to 8.2 (for $z=0.5$).

%%%%%%%%%%%%%%%%%%%%%%%%%%%%%%%%%%%%%%%
% \subsection{Query Evaluation on U-relations}
% \label{sec:exp-queries}

%%%!!!!!!!!!!!!!!!!!!!!!!!!!!!!!%%%%%%%%%%%%%%%%%%%%
% I cannot put this figure in longversion without compilation problems ?!
%%%!!!!!!!!!!!!!!!!!!!!!!!!!!!!!%%%%%%%%%%%%%%%%%%%%

\begin{figure}[h!]
\begin{tiny}
\begin{verbatim}
Merge Join  (cost=3187724.24..434887461.47 rows=14175759502 width=18)
 Merge Cond: (u_l_quantity.tid = u_l_extendedprice.tid)
 Join Filter: (((u_l_quantity.c1 <> u_l_extendedprice.c1) OR (
 u_l_quantity.w1 = u_l_extendedprice.w1)) AND 
 ((u_l_extendedprice.c1 <> u_l_discount.c1) OR 
 (u_l_extendedprice.w1 = u_l_discount.w1)) AND 
 ((u_l_extendedprice.c1 <> u_l_shipdate.c1) OR (u_l_extendedprice.w1 = u_l_shipdate.w1)))
 -> Merge Join  (cost=1381116.36..7243281.93 rows=224865665 width=79)
      Merge Cond: (u_l_shipdate.tid = u_l_quantity.tid)
      Join Filter: (((u_l_quantity.c1 <> u_l_shipdate.c1) OR 
      (u_l_quantity.w1 = u_l_shipdate.w1)) AND ((u_l_quantity.c1 <> u_l_discount.c1) 
      OR (u_l_quantity.w1 = u_l_discount.w1)))
       -> Merge Join  (cost=810344.64..1026829.84 rows=10650797 width=55)
           Merge Cond: (u_l_discount.tid = u_l_shipdate.tid)
           Join Filter: ((u_l_shipdate.c1 <> u_l_discount.c1) OR 
           (u_l_shipdate.w1 = u_l_discount.w1))
           -> Sort  (cost=269775.70..271512.42 rows=694689 width=31)
                Sort Key: u_l_discount.tid
                -> Seq Scan on u_l_discount  (cost=0.00..164374.00 rows=694689 width=31)
                     Filter: ((l_discount > '0.05') AND (l_discount < '0.08'))
           -> Sort  (cost=540568.94..545791.18 rows=2088896 width=24)
                Sort Key: u_l_shipdate.tid
                ->  Seq Scan on u_l_shipdate  (cost=0.00..171354.29 rows=2088896 width=24)
                     Filter: ((l_shipdate > '1994-01-01') AND (l_shipdate < '1996-01-01'))
       -> Sort  (cost=570771.73..576676.98 rows=2362101 width=24)
           Sort Key: u_l_quantity.tid
           -> Seq Scan on u_l_quantity  (cost=0.00..151169.98 rows=2362101 width=24)
               Filter: (l_quantity < '24')
 -> Sort  (cost=1806607.87..1824240.68 rows=7053122 width=35)
     Sort Key: u_l_extendedprice.tid
     ->  Seq Scan on u_l_extendedprice  (cost=0.00..136447.22 rows=7053122 width=35)
\end{verbatim}
\end{tiny}

\vspace{-1.5em}

\caption{Query plan for $Q_2$ ($\mathtt{s}=1,\mathtt{x}=0.1, \mathtt{z}=0.1$), as generated by PostgreSQL.}
\label{fig:qplan}
\vspace{-1em}
\end{figure}

\noindent{\bf Query Evaluation on U-relations.} We run four times our set of three queries on the 45
different datasets reported in Figure~\ref{fig:dbsize}. For each query
and correlation ratio, Figure~\ref{fig:perf} has a log-log scale
diagram showing the median evaluation (including storage) time in
seconds as a function of the scale and uncertainty parameters
\shortversion{(\cite{AKO07-Urel} also shows diagrams for
$z=0.25$)}. The different lines in each of the diagrams correspond to
different uncertainty ratios.

Figure~\ref{fig:perf} shows that the evaluation of our queries is
efficient and scalable. In our largest scenario, where the database
has size 13 GB and represents $10^{8\cdot 10^6}$ worlds with 1.4 GBs
each world, query $Q_3$ involving five joins is evaluated in less than
two and a half minutes. One explanation for the good performance is
the use of attribute-level representation. This allows to first
compute the joins locally using only the join attributes and later
merge in the remaining attributes of interest. Another important
reason for the efficiency is that due to the simplicity of our
rewritings, PostgreSQL optimizes the queries in a fairly good way.
\shortversion{(\cite{AKO07-Urel} shows an optimized query plan 
produced by the PostgreSQL `explain' statement for the rewriting of
$Q_2$.)}\longversion{ Figure~\ref{fig:qplan} shows an optimized query
plan produced by the PostgreSQL `explain' statement for the rewriting
of $Q_2$.}

The evaluation time varies linearly with all of our parameters. For
$Q_1$ ($Q_2$ and $Q_3$ respectively) we witnessed a factor of up to 6
(4 and 10 respectively) in the evaluation time when varying the
uncertainty ratio from 0.001 to 0.1. When the correlation ratio is
varied from 0.1 to 0.5, the evaluation time increases by a factor of
up to 3; this is also explained by the increase in the input and
answer sizes, cf.\@ Figures~\ref{fig:dbsize} and
\ref{fig:answersizes}. When the scale parameter is varied from 0.01 to
1, the evaluation time increases by a factor of up to 400; in case of
$Q_3$ and $z = 0.5$, we also noticed some outliers where the increase
factor is around 1000.  \longversion{ The considerably smaller
  evaluation time for $Q_3$ in case of scale 1, uncertainty 0.1, and
  correlation 0.25 occurs because for that scenario no `GERMANY' entry
  is generated for the nation table, thus the query answer is empty.}

{\noindent\bf  Effect of  attribute-level  representation.}   We  also
performed query evaluation on tuple-level U-relations, which represent
the  same     world-set as    the  attribute-level   U-relations    of
Figure~\ref{fig:dbsize}, and on Trio's ULDBs~\cite{BDSHW2006} obtained
by a (rather  direct) mapping from   the tuple-level U-relations.   To
date, Trio has no native support for  the poss operator or the removal
of  erroneous tuples in  the query answer, though  this  effect can be
obtained   as part  of  the confidence   computation\footnote{Personal
  communication  with  the  TRIO team  as  of June  2007.}.  For  that
reason, we decided to compare  the evaluation times of queries without
the poss  operator and without  the  (expensive) removal of  erroneous
tuples  or  confidence    computation  (which is  an  exponential-time
problem).   Since  our  data  exhibits  a   high degree  of  (randomly
generated) dependency, its  ULDB  representation has lineage and  thus
join  queries can introduce erroneous tuples  in  the answer. The Trio
prototype was set to use the (faster) SPI interface of PostgreSQL (and
not its default python implementation).

\begin{figure}[ht]
        \centering
        \includegraphics[scale=0.9]{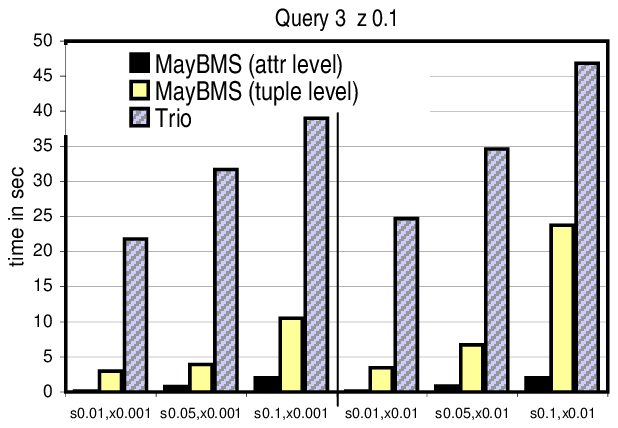}
        \vspace{-1em}
        \caption{Querying attribute- and tuple-level U-relations in MayBMS and ULDBs in
Trio.}
        \label{fig:comparison}
        \vspace{-1.5em}
\end{figure}

Figure~\ref{fig:comparison} compares the evaluation time on
attribute- and tuple-level U-relations in MayBMS, and ULDBs for
small scenarios of 1\% uncertainty, our lowest correlation factor 0.1,
and scale up to 0.1. On attribute-level U-relations, the queries
perform several times better than on tuple-level U-relations and by an
order of magnitude better than ULDBs. This is because attribute-level
data allows for late materialization: selections and joins can be
performed locally and tuple reconstruction is done only for successful
tuples.
% The worse timings for Trio seem to be caused by
% the fact that data is stored separately from its lineage and thus a query needs
% to access and create more than one relation.
We witnessed that an increase in any of our
parameters would create prohibitively large (exponential in the 
arity) tuple-level representations. For example, for scale 0.01 and
uncertainty 10\%, relation lineitem contains more than 15M tuples
compared to 80K in each of its vertical partitions.

%{\bf Column store awareness.} We experimented with early and late
%materialization of tuples by comparing the case where the merge
%operator is performed before the joins and the case where the merging
%is done as late as possible.
%%
%%, such that the joins on the key-values are considered first.
%%
%We observed that PostgreSQL executes both queries
%in approximately the same time, as for both cases the query optimizer
%produces similar execution plans, which use late materialization.

\vspace{-3mm}

%% file: bigtree.tex
\begin{footnotesize}
   \pstree[levelsep=.5cm,treesep=.2cm,nodesep=.05cm]{\TR{poss}}
   {
     \pstree{\TR{merge}}
     {
       \TR{$\pi_{\text{shippriority}}(o)$}
       \pstree{\TR{$\pi_{\text{o.orderkey,o.orderdate}}$}}
       {
         \pstree{\TR{$\Join_{\text{orderkey}}$}}
         {
           \pstree{\TR{merge}}
           {
              \TR{$\pi_{\text{orderkey}}(o)$}
              \pstree{\TR{$\pi_{\text{o.orderdate}}$}}
              {
                \pstree{\TR{$\Join_{\text{custkey}}$}}
                {
                  \pstree{\TR{merge}}
                  {
                    \TR{$\pi_{\text{custkey}}(o)$}
                    \pstree{\TR{$\sigma_{\text{o.orderdate}}$}}
                    {
                      \TR{$\pi_{\text{orderdate}}(o)$}
                    }
                  }
                  \pstree{\TR{merge}}
                  {
                    \TR{$\pi_{\text{custkey}}(c)$}
                    \pstree{\TR{$\pi_{\emptyset}$}}
                    {
                      \pstree{\TR{$\sigma_{\text{c.mktsegment}}$}}
                      {
                        \TR{$\pi_{\text{mktsegment}}(c)$}
                      }
                    }
                  }
                }
              }
           }          
           %\pstree{\TR{$U_{\text{l.orderkey}}$}}
           %{
             \pstree{\TR{merge}}
             {
               \TR{$\pi_{\text{orderkey}}(l)$}
               \pstree{\TR{$\pi_{\emptyset}$}}
               {
                 \pstree{\TR{$\sigma_{\text{l.shipdate}}$}}
                 {
                   \TR{$\pi_{\text{shipdate}}(l)$}
                 }
               }
             }
           %}
         }
       }
     }
   }
   \vspace{-0.3cm}
\end{footnotesize}

%%% Local Variables: 
%%% mode: latex
%%% TeX-master: t
%%% End: 

%% file: conclusion.tex
\section{Conclusion and Future Work}
\label{sec:conclusion}
\vspace*{-1em}

This paper introduces U-relational databases, a simple representation
system for uncertain data that combines the advantages of existing
systems, like ULDBs and WSDs, without sharing their drawbacks.
U-relations are exponentially more succinct than both WSDs and ULDBs.
Positive relational algebra queries are evaluated purely relationally
on U-relations, a property not shared by any other previous succinct
representation system.  Also, U-relations are a simple formalism which
poses a small burden on implementors.

We next briefly report on two current research directions.

{\noindent\bf Probabilistic U-relations.} U-relational databases can
be elegantly extended to model probabilistic information by just
adding a probability column $P$ to the world table $W$. For each
variable $x$, the sum of the values $\pi_P(\sigma_{\mathrm{Var}=x})(W)$
must equal one. We can then assign probability to any subset of the
world-set, described by a ws-descriptor $\overline{d}$, as the product
of probabilities of each variable assignment in $\overline{d}$.

The techniques for evaluating the operations of positive relational algebra
presented in this paper are applicable
in the probabilistic case without changes.
Computing the
confidences of the answer tuples is an inherently hard problem \cite{dalvi04efficient}. Our
current research investigates practical approximation techniques for
confidence computation.

{\noindent\bf Support for new language constructs.}
Following our
recent investigation on uncertainty-aware language constructs beyond
relational algebra~\cite{AKO07ISQL}, we identified common physical
operators needed to implement many primitives for the creation
and grouping of worlds.
%These operations are necessarily hard problems.
It appears that normalizing sets of ws-descriptors
in the sense of Section~\ref{sec:normalization}
plays an important
role in evaluating these operations and in confidence computation. We are
currently working on secondary-storage algorithms for normalization.

% The inlined representations of \cite{AKO07ISQL} are essentially
% tuple-level U-relational databases with a single component. Since we
% can always obtain this non-succinct form by composing all components
% into one, this implies that \cite{AKO07ISQL} provides us with
% effective although not yet efficient methods for evaluating the
% powerful query language introduced in that paper, world-set algebra,
% on U-relational databases.

%%% Local Variables: 
%%% mode: latex
%%% TeX-master: "icde2008_maybms"
%%% End: 